\documentclass[11pt,double]{article}
\usepackage[latin1]{inputenc}
\usepackage{amscd}
\usepackage{amsmath}
\usepackage{amssymb}
\usepackage{amsfonts}
\usepackage{latexsym}
\usepackage{amsthm}
\usepackage{mathrsfs}
\usepackage{booktabs}
\usepackage{caption}
\usepackage{booktabs}
\usepackage[nohead]{geometry}
\usepackage{epsfig}
\usepackage{algorithm}		
\usepackage{authblk}
\usepackage{cite}
\usepackage{helvet}         
\usepackage{courier}        
\usepackage{type1cm}        
\usepackage{url}
\usepackage{algpseudocode}
\usepackage{algpascal}
\usepackage{rotating}
\usepackage{multirow}
\usepackage[round]{natbib}
\usepackage{authblk}
\usepackage{bm}
\usepackage{bbm}
\usepackage{setspace}
\numberwithin{equation}{section}

\def\qmo{``}
\def\qmc{''}
\def\qmcsp{'' }

\def\by{\mathbf{y}} 

\def\0{\mbox{\bf{0}}}

\def\bZ{\mathbf{Z}}

\def\bc{\mathbf{c}} 
\def\bT{\mathbf{T}}

\def\bO{\mathbf{0}}

\def\balpha{\mbox{\boldmath $\alpha$}}

\def\bOmega{\mbox{\boldmath $\Omega$}}

\def\bpsi{\mbox{\boldmath $\psi$}}

\def\bnu{\mbox{\boldmath $\nu$}}

\def\bSigma{\mathbf{\Sigma}}
\def\btau{\mbox{\boldmath $\tau$}}

\def\sB{\mathsf{B}}
\def\sD{\mathsf{D}}
\def\sU{\mathsf{U}}
 
\def\sL{\mathsf{L}}

\def\sI{\mathsf{I}}

\def\sM{\mathsf{M}}
\def\sT{\mathsf{T}}
\def\sW{\mathsf{W}} 
 
\def\sR{\mathsf{R}}

\def\trasp{\mathsf{T}}

\newcommand{\invgamma}{\mathcal{IG}}

\def\qmo{``}
\def\qmc{''}
\def\qmcsp{'' }

\newcommand\bbone{\ensuremath{\mathbbm{1}}}

\newcounter{example}[section]
\def\theexample{\thesection.\arabic{example}}
{
\medskip\noindent\\
\refstepcounter{example} {\bf Example \theexample\hspace{2pt}}
-\hspace{2pt}}{%
 \begin{flushright}
 {\footnotesize $\square$}
 \end{flushright}
}

\title{\huge\bf Extracting the Italian output gap: a Bayesian approach}

\author[a]{M. Bernardi\footnote{Corresponding author: Sapienza University of Rome, Dept. MEMOTEF, Via del Castro Laurenziano, 9, 00161 ROME, \textrm{Tel.:} +39.06.49766972, \textrm{email address:} \textrm{mauro.bernardi@uniroma1.it.}}}
\author[b]{A. Di Ruggiero}

\affil[a]{MEMOTEF, Sapienza University of Rome}
\affil[b]{Economics and Law, University of Rome 2, Tor Vergata}

\begin{document}

\maketitle

\date{}

\begin{abstract}
\noindent During the last decades particular effort has been directed towards understanding and predicting the relevant state of the business cycle with the objective of decomposing permanent shocks from those having only a transitory impact on real output. This trend--cycle decomposition has a relevant impact on several economic and fiscal variables and constitutes by itself an important indicator for policy purposes. This paper deals with trend--cycle decomposition for the Italian economy having some interesting peculiarities which makes it attractive to analyse from both a statistic and an historical perspective. We propose an univariate model for the quarterly real GDP, subsequently extended to include the price dynamics through a Phillips curve. This study considers a series of the Italian quarterly real GDP recently released by OECD which includes both the 1960s and the recent global financial crisis of 2007--2008.
Parameters estimate as well as the signal extraction are performed within the Bayesian paradigm which effectively handles complex models where the parameters enter the log--likelihood function in a strongly nonlinear way. A new Adaptive Independent Metropolis--within--Gibbs sampler is then developed to efficiently simulate the parameters of the unobserved cycle. Our results suggest that inflation influences the Output Gap estimate, making the extracted Italian OG an important indicator of inflation pressures on the real side of the economy, as stated by the Phillips theory. Moreover, our estimate of the sequence of peaks and troughs of the Output Gap is in line with the OECD official dating of the Italian business cycle.\newline\newline
%
%
%
\vspace{0.25cm}\\
%
\end{abstract}


\section{Introduction}
\label{sec:intro}
%
%
The output gap (OG, henceforth) is the difference between the current level of output in the economy and the potential level that could be supplied without putting upward or downward pressure on inflation. For example Gali \citeyearpar{gali.2003} defines the output gap as: \qmo the deviation of output from its equilibrium level in the absence of nominal rigidities\qmc. In this context, the output gap is a summary indicator of the relative demand and supply components of economic activity. As such, the output gap measures the degree of inflation pressure in the economy and it is an important link between the real side of the economy -- which produces goods and services -- and the nominal side, price and its dynamic counterpart, e.g. inflation. All else equal, if the output gap is positive over time, so that actual output is greater than potential output, prices will begin to rise in response to demand pressure in key markets. Similarly, if actual output falls below potential output over time, prices will begin to fall to reflect weak demand.\newline
%
%
\indent Policymakers often use potential output to gauge inflation and typically define it as the level of output consistent with no pressure for prices to rise or fall. The use of business cycle indicators, like the OG, in monetary policy decisions reflects the judgement that demand--supply imbalances in particular markets, or the economy as a whole, provide indications about prospective inflationary pressures in product and labour markets. As argued by Walsh \citeyearpar{walsh.2003}, information about the future path of potential output is thus essential to evaluate whether the current stance of monetary policy is consistent with price stability.
If a central bank believes that the output or unemployment gap is positive when it is in fact negative, a consequent monetary policy tightening will tend to amplify the business cycle and intensify the downward pressure on inflation. As changes in the stance of monetary policy affect the aggregate economy only with a certain time lag, policymakers do not only require reliable estimates about the past or current state of the economy but also about its future evolution, see e.g. Koske and Pain \citeyearpar{koske_pain.2008}.
Moreover, indicators of the cyclical position of the economy are also important in the assessment of the sustainability of fiscal policy and the current fiscal stance, with actual budget positions being corrected for the impact of cyclical influences in order to gauge the underlying fiscal position. Otherwise, purely cyclical changes in the budget might falsely be treated as structural, potentially leading to serious policy mistakes. The level of the cyclically--adjusted budget balance, and hence the level of the output/unemployment gap matters when evaluating fiscal sustainability; the change in the gaps, is central to estimates of the current fiscal stance.\newline
%
%
\indent As we can not observe the level of potential output directly, even in retrospect, we need to estimate the size of the output gap from available indicators or from assumptions about the path of potential output. Estimating the output gap in real time is particularly difficult as changes in data may reflect movements in potential output, cyclical fluctuations, or both. Distinguishing between the underlying trend in potential output and cyclical fluctuations around it can be hard even once a much longer run of data becomes available.
%
%
%
%
%
This paper provides estimates of the potential capacity and the Output Gap for the Italian economy using unobserved component models for the period 1960--2013. The contributions of this paper are twofold. First, it fills the gap in the empirical literature by investigating the Italian business cycle for the period 1960--2013 using a series of quarterly Real GDP recently delivered by the OECD. Second, it is shown that employing an advanced econometric technique which combines structural unobserved component models with the Bayesian paradigm we are able to make effective and efficient inference on the crucial parameters concerning the cycle amplitude, phase and period.
%
%
%
%
%
%
%
Concerning the Italian OG extraction, the history of measurement of the Italian business cycle goes back to the seminal work of Gallegati and Stanca \citeyearpar{gallegati_stanca.1998} who analyse several macroeconomic series for the period 1861--1995 in order to provide some empirical regularities which help to understand the theoretical mechanisms underlying the business cycle fluctuations. Delli Gatti \textit{et al.} \citeyearpar{delli_gatti_etal.2005} analyse the nature and causes of business fluctuations in Italy over the \qmo long run\qmcsp emphasising the role of structural change in shaping each cyclical episode. From a methodological point of view, both those contributions rely on statistical filters such as the Hodrick--Prescott (\citeyear{hodrick_prescott.1997}, HP henceforth) and Baxter and King (\citeyear{baxter_king.1999}, BK henceforth) to extract a smoothed trend from an output series. Baffigi \textit{et al.} \citeyearpar{baffigi_etal.2013} provide a measure of the cyclical component of the real Italian Gross Domestic Product (GDP, henceforth) from 1861 to 2010 in order to assess the Phillips curve ability to explain the inflation dynamics. Their approach considers an autoregressive process for the inflation dynamics including additional predictors leading to the so--called generalised Phillips curve \textit{\`a la} Stock and Watson \citeyearpar{stock_watson.1999}. Among the exogenous variables they also include several predetermined measures of the Italian output gap and test the significance of the arising relationship. Their main finding suggests that the Phillips curve relation is verified only for the period following World War I.\newline 
\indent We contribute to the existing literature on the Italian Business cycle analysis in many aspects. Unlike Gallegati and Stanca \citeyearpar{gallegati_stanca.1998} and Delli Gatti \textit{et al.} \citeyearpar{delli_gatti_etal.2005} we propose structural unobserved components models for decomposing the output series into a permanent and a transitory component where the former identifies the potential capacity evolution while the latter refers to the OG pattern. Unobserved components models are useful tools for representing univariate and multivariate macroeconomic series as functions of latent processes having their own dynamics and represent a natural framework when dealing with the signal extraction problem. For an up to date and comprehensive introduction on structural time series models see e.g. Harvey \citeyearpar{harvey.1989} and Durbin and Koopman \citeyearpar{durbin_koopman.2012}. This setup is sufficiently general to cover structural time series models, also extended for some kind of nonlinearity, such as smooth transition (see e.g. Proietti, \citeyear{proietti.1998}), the ARIMA model based approach, the Beveridge--Nelson decomposition and structural VAR, and basically all parametric models and some non--linear extensions which however yield conditionally Gaussian models. Furthermore, the HP and BK filters as well as Christiano and Fitzgerald (\citeyear{christiano_fitzgerald.2003}, BP henceforth) Band--Pass approach are special cases of the aforementioned models which can be obtained by imposing specific parameters constraints, see e.g. Proietti \citeyearpar{proietti.2009} and Azevedo \textit{et al.} \citeyearpar{azevedo_etal.2006}. We firstly consider a univariate framework where the quarterly real GDP is decomposed into three orthogonal components: trend, cycle and irregular. Within this framework we compare several specifications differing for the stochastic properties of the underlying long--run component. As in Planas \textit{et al.} \citeyearpar{planas_etal.2008}, the cyclical component has a reduced--form ARMA(2,0) representation with complex roots to impose a pseudo--cyclical behaviour. Then we move to a bivariate model of output (real GDP) and inflation (CPI) embodying a Phillips--type relationship to link the inflation dynamics to the common unobserved cycle having the same stochastic representation as in the univariate case. Even in this case we compare several alternative specification for the long--run evolution of the real GDP series. A similar analysis has been previously undertaken by Zizza \citeyearpar{zizza.2006} and Bassanetti \textit{et al.} \citeyearpar{bassanetti_etal.2010} starting from the 80's. The main reason to derive estimates of the output gap employing a Phillips curve relationship is that inflation contains information about the gap. This conjecture has been empirical tested in several works for the US economy: see e.g. Kuttner \citeyearpar{kuttner.1994}, Harvey \citeyearpar{harvey.2011}, Harvey \textit{et al.} \citeyearpar{harvey_etal.2007}, Planas \textit{et al.} \citeyearpar{planas_etal.2008}.
Concerning the Italian economy instead, Baffigi \textit{et al.} \citeyearpar{baffigi_etal.2013} concludes for statistical validity of the curve which essentially means that for the Italian case nominal rigidities are not so strength to prevents the adjustment of the output through the long-run path when and endogenous shock affects the economy. If this assumption is valid the prices dynamics is able to display all its power in the explanation of the OG measure. However, this latter work only consider annual data for the period 1860--2010 which consists of different inflationary regimes. In this paper we confine our analysis to the period post--II World War and we use quarterly observations for both the CPI and the Real GDP recently released by OECD.\newline
\indent The second major contribution concerns the employed estimation methodology. In particular, we estimate model parameters within the Bayesian paradigm. Bayesian inference arises quite naturally as the standard tool when dealing with latent variable models that can be cast in state space, see e.g. Fr\"uhwirth-Schnatter \citeyearpar{fruhwirth-schnatter.2006} and Capp\'e \textit{et al.} \citeyearpar{cappe_etal.2005}. However, despite the simple structure of some of the models here considered, standard Markov chain Monte Carlo (MCMC) algorithms such as the Gibbs sampling and the Metropolis--Hastings, are not suitable for trend--cycle decomposition and fail to converge even after many iterations. The main reason for the very slow convergence and extremely high autocorrelation usually displayed by standard MCMC methods relies on how the cycle parameters enter the likelihood function which becomes highly nonlinear with respect to those parameters. 
A proposal distribution which better approximates the posterior distribution of the model parameters is therefore desirable, and we provide an automatic approach to constructing such distributions.
We introduce an adaptive--within--Gibbs sampling algorithm where the parameters of the proposal distributions are continuously updated to tailor the shape of the proposal distribution to that of the target. The distinctive characteristic of adaptive algorithms with respect to standard MCMC methods is the presence of a proposal distribution whose parameters are modified during the simulation process to minimise a distance with the target distribution using the past iterations of the sampler. 
Adaptive MCMC methods are now well developed to simulate form complex and usually intractable posterior distributions and their theoretical properties are well understood, see e.g. Andrieu and Thoms \citeyearpar{andrieu_thoms.2008} and Liang \textit{et al.} \citeyearpar{liang_etal.2010}, Atchad\'e and Rosenthal \citeyearpar{atchade_rosenthal.2005},
Atchad\'e \textit{et al.} \citeyearpar{atchade_etal.2011}. The convergence and ergodicity of these Adaptive--MCMC has been investigated by Roberts and Rosenthal \citeyearpar{roberts_rosenthal.2007}.
The general framework for designing our adaptive--MCMC algorithm is built around the Independent Metropolis--withing--Gibbs combined with the data augmentation approach of Tanner and Wong \citeyearpar{tanner_wong.1987}. We propose to augment the posterior distribution by the latent factors, and to simulate from the resulting joint distribution by using a two--step Gibbs sampler. In the first step we simulate the latent states conditional on model parameters using a Forward--Filtering--Backward--Smoothing (FFBS) algorithm, see e.g. Fr\"uhwirth--Schnatter \citeyearpar{fruhwirth-schnatter.2006} and Durbin and Koopman \citeyearpar{durbin_koopman.2012}. In the second step, we simulate from the full conditional of the static parameters using a blocked Independent Metropolis proposal for those parameters whose conditional distribution is not known. For the remaining parameters standard Gibbs moves are proposed. The proposal parameters are then adapted using the past draws of the sampler by generalising the strategy proposed in Andrieu and Thoms \citeyearpar{andrieu_thoms.2008} (Algorithm 2). For each block, we minimise the Kullback--Leibler divergence between the proposal scheme and the correspondent full conditional, using the stochastic approximation algorithm of Robbins and Monro \citeyearpar{robbins_monro.1951}.
This sampling method enables efficient mixing of the resulting chain and easily adapts to the different model specifications we consider. The convergence and ergodicity of these algorithms is guaranteed by a careful design of the adaptation strategy, see Andrieu \textit{et al.} \citeyearpar{andrieu_moulines.2006}.\newline
%
%
\indent The paper is organised as follows. In Section \ref{sec:literature_review} we present a brief review of the related literature. In Section \ref{sec:og_models} we lay out the models we propose to analyse the Italian OG. Section \ref{sec:prior_specification} specifies the prior distributions for the model parameters which are necessary to perform Bayesian inference while Section \ref{sec:og_inference} details the Metropolis--within--Gibbs algorithm used to perform the parameters estimation. In Section \ref{sec:application} describe the Italian data, present our primary results and discuss the main  implications  of the extracted cycles with respect to the existing italian business cycle datings. We summarise and conclude in Section \ref{sec:conclusion}.
\section{A brief literature review on output gap}
\label{sec:literature_review}
%
\noindent This paper deals with the signal extraction of the Italian Output Gap. Over the last decades the literature on business cycle estimation has grown rapidly mainly because of its crucial role for economic analysis and for the consequent policy implications. As an example, a business cycle indicator, such as the output gap, providing a signal of inflationary pressure in the economy, may help in defining the appropriate monetary policy. Moreover, it could represent a valid instrument to address the question on to what extent unemployment can be attributed to a lack of overall demand (see e.g. Solow, \citeyear{solow.2000}).\newline
\indent However, despite the fact that OG measures are often taken for granted by macroeconomy analysts, the issue regarding its measure is quite controversial and it is indisputably related to very different concepts and theories behind it. Whenever our aim is to employ OG as representative of cyclical fluctuations in the economy, we can build several statistical models in which OG represents the stationary cyclical component of overall output.\newline
\indent The literature regarding the statistical, or theory--free, methods to estimate OG is mainly based on filtering techniques and it  has in the work of Hodrick and Prescott (\citeyear{hodrick_prescott.1997}, HP henceforth) and Baxter and King (\citeyear{baxter_king.1999}, BK henceforth) its seminal contributions. The HP filter is the most common filter used to decompose economic time series thanks to its simple calculation and implementation in many econometric software packages. Nevertheless, it presents several shortcomings since OG turns out to be a simple white noise component rather than having a cyclical persistent dynamic pattern. Moreover, it relies on a integrated random walk dynamics for the trend component which may be unreliable for most economic time series. The misspecification of the cyclical component as well as the lack of enough flexibility in the long run dynamics represent severe limitations of these kind of filters that also affect the inferential aspect. In fact, another crucial shortcoming of the HP filter is the arbitrary choice of the smoothing parameter. Finally, a relevant problem, common to other filters like the band pass filter of Baxter and King \citeyearpar{baxter_king.1999}, is the so--called end--of--sample bias which prevents us from extracting the cyclical signal for the most important part of the series. Within this framework the estimates of the cycle at extremes of the available sample become less credible. To overcome this problem, Christiano and Fitzgerald (\citeyear{christiano_fitzgerald.2003}, BP henceforth) provide a model--based structural parametric interpretation of the band pass filters embedding their estimation into the Wiener--Kolmogorov optimal signal extraction theory.\newline  
\indent The unobserved components (UC) methods treat both long term dynamics of the series (potential output) and cyclical component (output gap, OG) as unobserved variables, employing statistical techniques to decompose a time series into these, usually orthogonal components. However, unlike the aforementioned filter techniques, in which basically the nature of the trend is set a priori, the UC methods are based on stochastic de--trending tools which are picked up by firstly analysing  the properties of the time series. These methods rely on a precise specification of the process generating both cycle and the trend. In particular, the OG can be seen as the transitory component in the GDP series and it is usually specified as an autoregressive and moving average (ARMA) process with complex roots to introduce a pseudo--cyclical dynamics. The long run dynamics, i.e. the potential output, is usually specified as integrated process of different order. All these models can be written in state space form and can be analysed by using Kalman filter, (see e.g. Harvey, \citeyear{harvey.1989} and Durbin and Koopman, \citeyear{durbin_koopman.2012}). Within this general framework, the most relevant contributions are those of Clark \citeyearpar{clark.1987}, Harvey and J\"aeger \citeyearpar{harvey_jaeger.1993} which replace the misspecified irregular component in the HP filter by a stationary stochastic cycle having an autoregressive polynomial of order 2 with complex conjugate roots and a moving average error term. Harvey and Trimbur \citeyearpar{harvey_trimbur.2003} further extend the model specification of Harvey and J\"aeger \citeyearpar{harvey_jaeger.1993}, by proposing a general class of model based filters for extracting trend and cycles in macroeconomic time series, showing that the design of low--pass and band--pass filters can be considered as a signal extraction problem in an unobserved components framework.\newline
\indent The UC methods have subsequently been extended into a multivariate approach, considering besides GDP other macroeconomic series which can provide relevant information regarding the business cycle. The multivariate approach can be considered much more theoretically grounded with respect to the univariate one, since it exploits common relationships in macroeconomics, like for example the Phillips curve. In this respect, Kuttner \citeyearpar{kuttner.1994} analyses the joint behaviour of the real GDP and CPI inflation in order to back out an estimation of potential GDP and OG employing the maximum likelihood (ML) approach. Planas {\it et al.} \citeyearpar{planas_etal.2008} instead estimate the Kuttner \citeyearpar{kuttner.1994}'s model within a Bayesian framework for both the US and EU macro area. In their model specification, Planas {\it et al.} \citeyearpar{planas_etal.2008} re--parameterize the cycle as an AR(2) directly imposing the complexity of the AR polynomial, which also facilitates the prior elicitation process. Another relevant contribution is that of Harvey {\it et al.} \citeyearpar{harvey_etal.2007} which also consider the Bayesian estimation of a bivariate model of output and inflation, where the cycle in inflation is driven by the output gap plus an idiosyncratic cycle. The main difference with the Planas \textit{et al.} \citeyearpar{planas_etal.2008} model is represented by the cycle specification having an ARMA(2,1) reduced form representation.\newline
\indent Regarding the theory--based approaches we can divide them in two macro categories. Ones which are based on the structural VAR analysis, the other which relies upon the aggregate production function. The latter is very difficult to implement since it requires a collection of data which are not easily obtainable such as the stock of capital and the non--accelerating inflation rate of unemployment (NAIRU). The production function approach has been considered in Proietti {\it et al.} \citeyearpar{proietti_etal.2007} to estimate the business cycle in the Euro area. Output gap estimates implied by structural VAR models have been theoretically investigated in Mitchell {\it et al.} \citeyearpar{mitchell_etal.2008} and applied to the empirical investigation of the European business cycle. Both these approach have potential benefits but also display several limitations. Their main benefit consists on combining information coming from different time series. However, as argued by Proietti {\it et al.} \citeyearpar{proietti_etal.2007}, the production function approach is a method that provides ad hoc solutions to the problem of extracting a cyclical signal because the provided OG estimates strongly depend on the mathematical form of the chosen production function. The structural VAR approach instead is a little bit restrictive in the specification of the form of the cyclical factor common to all the considered series. Moreover, in conveying information coming from different sources particular attention should be devoted to the problem of selecting series having the same phase, period and amplitude.\newline
%
%
\indent Concerning the italian case, several attempts have already been made in order to extract consistent estimates of the Output Gap using different structural and reduced form models or non--parametric approaches like the HP, BK filters. However, only a few attempts have been made to extract the Italian OG using quarterly data, mainly for the period starting form 1980, in which official estimates of quarterly Italian real GDP begun to be supplied. From a classical perspective, the most recent contribution is that of Zizza \citeyearpar{zizza.2006} which estimate the Italian output gap for the period 1980--2004 using quarterly data. In particular, Zizza \citeyearpar{zizza.2006} employs several univariate, (real GDP), bivariate (real GDP, CPI inflation) and multivariate (real GDP, CPI inflation and industrial production, or real GDP, CPI inflation and unemployment) structural unobserved component models. Parameters are estimated within the classical paradigm by maximising the likelihood provided by the Kalman filter (see Durbin and Koopman \citeyear{durbin_koopman.2012} for a comprehensive and up to date introduction to the inference on state space models). The obtained estimated value for the period of the italian OG ranges from 3 years (for the bivariate models) to 5 years (for the univariate models), while the estimated cycle amplitude, as a percentage of trend GDP, ranges from 0.12 (for  the univariate model) to 0.30 (for the bivariate models). Busetti and Caivano \citeyearpar{busetti_caivano.2013} instead  deal with the trend--cycle decomposition of the Italian output within a Bayesian structural framework where real GDP and Inflation share the same unobserved cycle modelled as a structural component as in Harvey \textit{et al.} \citeyearpar{harvey_etal.2007}. Their result strongly differ from those obtained by Zizza \citeyearpar{zizza.2006}, that consider about the same period (1980--2013), both in terms of cycle phase and amplitude, for about the same period.\newline
\indent Bassanetti {\it et al.} \citeyearpar{bassanetti_etal.2010} exploit four different univariate approaches: a Bayesian unobserved component model, an univariate autoregressive model, a production function approach and a structural VAR approach, and combine them using the posterior model probabilities using a sort of Bayesian model averaging technique. The generated OGs happen to provide a proper description of the Italian business cycle after being compared with the official OG quarterly measures published by the Organisation for Economic Co--operation and Development (OECD, henceforth). Bayesian estimates are obtained imposing informative prior distributions on the cycle amplitude and period: specifically, the amplitude parameter follows a uniform distribution and must be larger than 0.5 and smaller than 1, while the cycle period is assumed to be distributed according to a Beta with parameters $\alpha=2$ and $\beta=5.1$, which implies a mode corresponding to a cycle of 32 quarters. In addition, the cycle period is restricted in order to exclude too high and too low frequencies, up to 12 and over 48 quarters, respectively.\newline
\indent Gallegati and Stanca \citeyearpar{gallegati_stanca.1998}, employs Hodrick Prescott filter to extract a quarterly level cycle indicator for the period 1960--1995. The estimated values for the cycle period is 5.2 years while the average amplitude is 0.12. Comparing those results with the ones of the other G7 countries it turns out that the amplitude and the period of the Italian output fluctuations are much more higher with respect to them.\newline
\indent At annual level, the efforts were mainly addressed to provide a business cycle dating which could help to identify the major turning points of the Italian economy.
%
\section{Output gap models}
\label{sec:og_models}
%
\noindent In this section we introduce our basic framework to extract the Output Gap for Italy. In particular univariate structural models for the real GDP are considered in Section \ref{sec:univ_og_models} and subsequently extended to the bivariate case (real GDP, Inflation) in Section \ref{sec:biv_og_models}.
%
\subsection{Univariate models}
\label{sec:univ_og_models}
%
Univariate time series models for GDP have been previously considered by Zizza \citeyearpar{zizza.2006}, and Bassanetti \textit{et al.} \citeyearpar{bassanetti_etal.2010}, while Delli Gatti \textit{et al.} \citeyearpar{delli_gatti_etal.2005} extract the cyclical signal in frequency domain using low--pass filters like the Baxter and King \citeyearpar{baxter_king.1999}. For a comprehensive and up to date overview of the frequency domain methods to extract the Italian output gap we refer the reader to the book of Gallegati and Stanca \citeyearpar{gallegati_stanca.1998}.\newline
\indent In this section we introduce the basic univariate framework for the extraction of the cyclical signal from the quarterly Italian real GDP adjusted to account for seasonal cycles, denoted by $y_t$. As mentioned in the Introduction, our approach acknowledges that neither the potential output nor the output gap is directly observed, and to treat each as a state variable in a state space representation, rather than to simply replace either with measured proxies. The payoff from this approach comes from using the Kalman filter and smoothing techniques to extract the estimated output gap implied by the behaviour of real GDP. In particular, the general model specification decomposes the observed series $y_t$ into orthogonal components 
\begin{equation}
y_t=\mu_t+\psi_t+\epsilon_t,\qquad\epsilon_t\sim\mathcal{N}\left(0,\sigma_\epsilon^2\right),
\label{eq:univ_og_meas}
\end{equation}
where $\mu_t$ is the trend component, i.e. a slowly varying component $\mu_t$ accounting for the long run evolution of the real GDP, $\psi_t$ is the cyclical transitory component accounting for the deviations  
of the real production from its equilibrium path, and $\epsilon_t$ is an erratic component which is included with the aim of extracting a less erratic signal from the data. This erratic unpredictable component is quite important because it accounts for short--run demand shocks affecting the Italian GDP at quarterly frequency and typically vanish within a year. The benefits for including such a stochastic component in the trend--cycle decomposition of Italian GDP goes far beyond the issue of a correct specification of the model DGP since it affects also the resulting signal extraction process. In particular, the resulting estimated cyclical component, i.e. the output gap, will be less variable in its amplitude. We assume the erratic component $\epsilon_t$ is normally distributed with homoscedastic variance. Summarising, the real GDP is linearly projected on unobservable variables, the potential output $\mu_t$ and the output gap $\psi_t$ both of them having their own dynamics. The trend $\mu_t$ reflects the impact of permanents shocks on the equilibrium level of output; and the output gap $\psi_t$, is the stationary component of the output associated with nominal rigidities in the economy.\newline
\indent Concerning the specification of the unobservable components dynamics $\left(\mu_t,\psi_t\right)$, we assume a local trend model (LT, henceforth) for the long--run component $\mu_t$ and an autoregressive component of order two for the cyclical deviations $\psi_t$, i.e.
\begin{eqnarray}
\begin{array}{lc}
\mu_{t}=\mu_{t-1}+\beta_{t-1}+\eta_t, & \qquad\eta_t\sim\mathcal{N}\left(0,\sigma^2_{\eta}\right)\\
\beta_{t}=\beta_{t-1}+\zeta_t,& \qquad\zeta_t\sim\mathcal{N}\left(0,\sigma^2_{\zeta}\right),\\
\end{array}
\label{eq:univ_og_model_trend_dyn}
\end{eqnarray}
where the term $\beta_t$ is included in the specification to ensure that the resulting process for $\mu_t$ will be integrated of order two. The form of the transition equations \eqref{eq:univ_og_model_trend_dyn} with both $\sigma_\eta^2$ and $\sigma_\zeta^2$ greater than zero, allows the trend level and the trend slope to vary over time. We call this specification $\mathcal{M}_{\sU}^{\sL\sT}$ where the subscript \qmo$\sU$\qmcsp stands for \qmo Univariate\qmcsp while the superscript \qmo$\sL\sT$\qmcsp stands for \qmo Local Trend\qmc, and \qmo$\mathcal{M}$\qmcsp denotes the considered model specification. Of course, this model specification nests different dynamics for the trend component. For example, fixing the variance of the slope component $\beta_t$, i.e. $\sigma_\zeta^2$ to be zero, results in a Local Level with Drift model (LLD, henceforth), which corresponds to a integrated process of order one with deterministic drift for $\mu_t$. We call this model specification $\mathcal{M}_\sU^{\sL\sL\sD}$. The Local Level model (LL, henceforth) corresponds instead to a specification of the univariate OG equations \eqref{eq:univ_og_model_trend_dyn} where the dynamics for the slope $\beta_t$ is totally absent. We denote this model with the label $\mathcal{M}_{\sU}^{\sL\sL}$. Interestingly, the state equation specification corresponding to the structural model representation of the HP filter requires the variance of the trend $\sigma_\eta^2$ to be fixed to zero, resulting in a smoother trend estimate. We call this model specification $\mathcal{M}_\sU^{\sI\sR\sW}$. The reduced form of the trend dynamics $\Delta^2\mu_t=\zeta_t$ corresponds to the integrated random walk model of Young \textit{et al.} \citeyearpar{young_etal.1991}. The HP filter simply considers this specification $\Delta^2\mu_t=\zeta_t$ for the observable variable $y_t$ in a regression where this term is penalised against overfitting, see e.g. Kaiser and Maravall \citeyearpar{kaiser_maravall.2005}.\newline 
\indent Completing the specification of the state variables, $\psi_t$ is assumed to be a stationary autoregressive process of order 2, following Harvey \citeyearpar{harvey.1985}, Watson \citeyearpar{watson.1986}, Clark \citeyearpar{clark.1987}, Harvey and J\"aeger \citeyearpar{harvey_jaeger.1993}
\begin{eqnarray}
\psi_t =\phi_1\psi_{t-1}+\phi_2\psi_{t-2}+\kappa_t,& \kappa_t\sim\mathcal{N}\left(0,\sigma^2_{\kappa}\right),
\label{eq:univ_og_model_cycle_dyn}
\end{eqnarray}
where $\left(\phi_1,\phi_2\right)$ are the autoregressive parameters and $\kappa_t$ is the Gaussian innovation term. The autoregressive component dynamics for the output gap variable $\psi_t$ is restricted to have complex roots to be effective in modelling pseudo--cyclical behaviours, see e.g. Box \textit{et al.} \citeyearpar{box_etal.2008}. To impose the stationarity and pseudo--cyclical behaviour we consider the approach of Planas \textit{et al.} \citeyearpar{planas_etal.2008} and reparameterise the autoregressive parameters $\left(\phi_1,\phi_2\right)$ in terms of the amplitude $\rho$ and the frequency $\lambda$ of the resulting cycle, i.e. $\phi_2=-\rho^2$ and $\phi_1=\frac{2\pi}{\lambda}$. The reduced form of the model is an ARIMA(2,2,0) with constrained parameters for identifiability pourpouses. An alternative structural representation of the cycle has been consider by Harvey \textit{et al.} \citeyearpar{harvey_etal.2007} for the trend--cycle decomposition of US economy using multiple time series. The major difference with the model here considered relies in the cyclical component which is modelled using a structural bivariate process having an ARMA(2,1) reduced form representation, see also Harvey \citeyearpar{harvey.1989}, Durbin and Koopman \citeyearpar{durbin_koopman.2012} and Proietti \citeyearpar{proietti.2009}. In principle any ARMA(2,q) process with $q\geq0$ provide the same stochastic cycle, the only difference being in the persistence of the impact of lagged endogenous deviations from the estimated cycle (MA components).\newline 
\indent Gathering the measurement equation \eqref{eq:univ_og_meas}, and the transition equations \eqref{eq:univ_og_model_trend_dyn}--\eqref{eq:univ_og_model_cycle_dyn} together, we get the following general state space representation for the univariate models
\begin{align}
y_t & =\mu_t+\psi_t+\epsilon_t, & \epsilon_t\sim\mathcal{N}\left(0,\sigma^2_{\epsilon}\right)
\label{eq:univ_og_model_state_space_form_1}\\
\mu_{t}& =\mu_{t-1}+\beta_{t-1}+\eta_t, & \eta_t\sim\mathcal{N}\left(0,\sigma^2_{\eta}\right)
\label{eq:univ_og_model_state_space_form_2}\\
\beta_{t}& =\beta_{t-1}+\zeta_t,& \zeta_t\sim\mathcal{N}\left(0,\sigma^2_{\zeta}\right)
\label{eq:univ_og_model_state_space_form_3}\\
\psi_t& =\phi_1\psi_{t-1}+\phi_2\psi_{t-2}+\kappa_t,& \kappa_t\sim\mathcal{N}\left(0,\sigma^2_{\kappa}\right)
\label{eq:univ_og_model_state_space_form_4}
\end{align}
where $\Theta=\left(\phi_1,\phi_2,\sigma_\epsilon,\sigma^2_\eta,\sigma^2_\zeta,\sigma^2_\kappa,\sigma^2_\epsilon,\sigma^2_\kappa\right)$ is a vector of model parameters. The model defined in equations \eqref{eq:univ_og_model_state_space_form_1}--\eqref{eq:univ_og_model_state_space_form_4} admits several different specification as special cases. In particular, setting the variance of the slope component $\sigma_\zeta^2$ to zero results in a local trend model with deterministic drift, while setting the variance of the measurement equation $\sigma_\epsilon^2$ to zero, the resulting model specification delivers an exact decomposition of the observed signal in trend and cyclical components.
%
\subsection{Bivariate models}
\label{sec:biv_og_models}
%
In this section we introduce our general bivariate framework along. In particular, we consider a bivariate OG model for quarterly Italian real GDP ($y_t$) and quarterly rate of inflation ($\pi_t=\Delta p_t$), where $p_t$ is the logarithm of the Italian quarterly CPI. In particular, we augment the measurement equation of the univariate models defined in Section \ref{sec:univ_og_models} by a second equation accounting for the inflation dynamics, in the following way
\begin{eqnarray}
\pi_t =\tau_t+\varepsilon_t,\qquad\varepsilon_t\sim\mathcal{N}\left(0,\sigma^2_{\varepsilon}\right),
\label{eq:biv_og_model_infl_meas}
\end{eqnarray}
where $\tau_t$ is an unobserved component describing the long--run behaviour of the $\pi_t$ series and $\varepsilon_t$ is an idiosyncratic error term, so that the inflation equation is decomposed into a core inflation $\tau_t$ plus a transitory component. Following the seminal work of Kuttner \citeyearpar{kuttner.1994}, the changes in the core inflation are driven by the output gap through a Phillips--type relation
\begin{equation}
\Delta\tau_t =\theta_{\psi}\left(L\right)\psi_t+\xi_t,\qquad\xi_t\sim\mathcal{N}\left(0,\sigma^2_{\xi}\right),
\label{eq:biv_og_model_core_infl_dyn}
\end{equation}
where $\Delta\tau_t=\tau_t-\tau_{t-1}$ denotes the first difference of the inflation, $\theta_{\psi}\left(L\right)=\theta_0+\theta_1L+\ldots+\theta_pL^p$ is a polynomial of order $p$ in the lag operator $L$ and $\xi_t$ is an idiosyncratic component. This specification reflects a backward--looking inflation expectation idea and is supported by Furher and Moore \citeyearpar{furher_moore.1995}, who argued for the presence of substantial backward--looking behaviour in estimating the inflation equation. The specification of equation \eqref{eq:biv_og_model_core_infl_dyn} is substantially differs from that proposed by Planas \textit{et al.} \citeyearpar{planas_etal.2008} who included also a term accounting for the first difference of the lagged output $\left(\Delta y_{t-1}\right)$ into the equation specifying the dynamics for the first difference of the inflation. Concerning this aspect, it is important to recognise that the Planas \textit{et al.} \citeyearpar{planas_etal.2008} specification is a little bit more general than that considered here in the sense that it correspond exactly to the Phillips' curve idea. Moreover, the significance of the inclusion of the term $\Delta y_{t-1}$ can be tested a posteriori by performing a statistical test on the correspondent loading parameter. Nevertheless, we observe also that Planas \textit{et al.} \citeyearpar{planas_etal.2008} assume a process integrated of order one for the real GDP dynamics $\left(y_t\right)$ while we allow also for an integrated of order 2 process for $y_t$. In the general framework considered here, where different specifications for the real output dynamics are tested, it becomes quite difficult to include such a term, preserving the stationarity of the resulting inflation dynamics. In addiction, the relevance of the inclusion of the equation for the inflation dynamics in the model specification of the OG equation in the specific case of Italy may be obscured by the presence of the lagged output. This is much more evident in the light of the fact that we compare different model specification in which the term $y_{t}$ is not always first--difference stationary and should be differentiated twice to become stationary. The different order of integration of $y_t$ which is necessary for its inclusion as predetermined regressor in the inflation dynamics probably will result at the end in an heterogeneous impact over the inflation contribution across model leading to meaningless results. Following the univariate approach, we consider the following specification for the real output dynamics: $\mathcal{M}_{\sB}^{\sL\sT}$ for the bivariate local trend model, $\mathcal{M}_\sB^{\sL\sL\sD}$ for the local level with drift model, $\mathcal{M}_{\sB}^{\sL\sL}$ for the local level model and $\mathcal{M}_\sB^{\sI\sR\sW}$ for the model--based HP filter extended to account for the cyclical behaviour of the output deviations from the long--term path.\newline
\indent Concerning the autoregressive polynomial $\theta_{\psi}\left(L\right)$ that loads the cyclical deviations into the inflation dynamics, following Kuttner \citeyearpar{kuttner.1994}, we consider an autoregressive of order one specification, i.e. $\theta_{\psi}\left(L\right)=\theta_0+\theta_1L$. Gathering the measurement equations \eqref{eq:univ_og_meas}--\eqref{eq:biv_og_model_infl_meas}, and the transition equations \eqref{eq:univ_og_model_trend_dyn}--\eqref{eq:univ_og_model_cycle_dyn}--\eqref{eq:biv_og_model_core_infl_dyn} together, we get the following general state space representation for the bivariate models
\begin{align}
y_t& =\mu_t+\psi_t+\epsilon_t,& \epsilon_t\sim\mathcal{N}\left(0,\sigma^2_{\epsilon}\right)
\label{eq:biv_og_model_specification_1}\\
\mu_{t}& =\mu_{t-1}+\beta_{t-1}+\eta_t, & \eta_t\sim\mathcal{N}\left(0,\sigma^2_{\eta}\right)
\label{eq:biv_og_model_specification_2}\\
\beta_{t}& =\beta_{t-1}+\zeta_t,&\zeta_t\sim\mathcal{N}\left(0,\sigma^2_{\zeta}\right)
\label{eq:biv_og_model_specification_3}\\
\psi_t& =\phi_1\psi_{t-1}+\phi_2\psi_{t-2}+\kappa_t,&\kappa_t\sim\mathcal{N}\left(0,\sigma^2_{\kappa}\right)
\label{eq:biv_og_model_specification_4}\\
\pi_t& =\tau_t+\varepsilon_t,&\varepsilon_t\sim\mathcal{N}\left(0,\sigma^2_{\varepsilon}\right)
\label{eq:biv_og_model_specification_5}\\
\tau_t& =\tau_{t-1}+\theta_0\psi_t+\theta_1\psi_{t-1}+\xi_t,&\xi_t\sim\mathcal{N}\left(0,\sigma^2_{\xi}\right)
\label{eq:biv_og_model_specification_6}
\end{align}
where $\Xi=\left(\phi_1,\phi_2,\theta_0,\theta_1,\sigma^2_\eta,\sigma^2_\zeta,\sigma^2_\kappa,\sigma_\varepsilon^2,\sigma^2_\epsilon,\sigma^2_\xi\right)$ is a vector of model parameters.
%
%
\subsection{State space representation}
\label{sec:og_models_ss_representation}
%
In this section we detail the state space representation of the OG models defined in the previous sections. In particular, we refer to the bivariate specification of the output gap models defined in Section \ref{sec:biv_og_models}. The Gaussian state space form we refer to in this work is defined in Durbin and Koopman \citeyearpar{durbin_koopman.2012}
\begin{align}
\by_t & =\bc+\bZ\balpha_t+\boldsymbol{\epsilon}_t, & \boldsymbol{\epsilon}_t\sim\mathcal{N}_d\left(\bO,\bSigma\right)\\
\balpha_{t+1}& =\mathbf{d}+\mathbf{T}\balpha_t+\boldsymbol{\eta}_t, & \boldsymbol{\eta}_t\sim\mathcal{N}_p\left(\bO,\bOmega\right)\\
\boldsymbol{\alpha}_1&\sim \mathcal{N}_p\left(0,\varkappa\mathbb{I}_p\right),
\end{align}
where $\by_t$ is a $d$--dimensional vector of observations at time $t$, $\bc$ and $\mathbf{d}$ are vectors of dimension $d$ and $p$ respectively, of constant terms,  is a vector of dimension $p$ of unobservable variables, $\bZ$ is a matrix of dimension $\left(d\times p\right)$ of loadings parameters, $\bT$ is the transition matrix of dimension $\left(p\times p\right)$, and $\left(\boldsymbol{\epsilon}_t,\boldsymbol{\eta}_t\right)$ are orthogonal vectors, of dimension $d$ and $p$ respectively, of Gaussian innovations with positive definite variance--covariance matrices $\bSigma$, $\bOmega$. To complete the state space definition we need to specify a distribution for the initial vector of latent states $\balpha_1$ which is Gaussian with mean equal to zero and diagonal variance--covariance matrix proportional to $\varkappa$ large enough to ensure a diffuse initialisation of the states. For an extensive discussion on the initialisation of the Kalman filter, see Durbin and Koopman \citeyearpar{durbin_koopman.2012}.\newline
\indent To cast the bivariate OG model defined in equations \eqref{eq:biv_og_model_specification_1}--\eqref{eq:biv_og_model_specification_6} in state space form, we need to specify the vector of observations $\by_t=\left(y_t,\pi_t\right)^\trasp$ with 
$d=\dim(\by_t)=2$, the vector of latent variables $\balpha_t=\left(\mu_t,\beta_t,\psi_t,\psi_{t-1},\tau_t\right)$, with $p=\dim(\balpha_t)$, and the state space relevant matrices $\left(\bc,\mathbf{d},\bZ,\bT,\bSigma,\bOmega\right)$
\begin{align}
\bZ&=\left[\begin{array}{ccccc}
1 & 0 & 1 & 0 & 0\\
0 & 0 & 0 & 0 & 1\\
\end{array}\right]
\label{eq:biv_og_model_general_ssf_mZ}\\
\bSigma&=\left[\begin{array}{cc}
\sigma_\epsilon^2 & 0\\
0 & \sigma^2_\varepsilon\\
\end{array}\right]
\label{eq:biv_og_model_general_ssf_mSigma}\\
\bT&=\left[\begin{array}{ccccc}
1 & 1 & 0 & 0 & 0\\
0 & 1 & 0 & 0 & 0\\
0 & 0 & \phi_1 & \phi_2 & 0 \\
0 & 0 & 1 & 0 & 0\\
0 & 0 & \theta_0\phi_1 + \theta_1 & \theta_0\phi_2 & 1\\ 
\end{array}\right]
\label{eq:biv_og_model_general_ssf_mT}\\
\bOmega&=\left[\begin{array}{ccccc}
\sigma^2_\eta & 0 & 0 & 0 & 0\\
0 & \sigma^2_\zeta & 0 & 0 & 0\\
0 & 0 & \sigma^2_\kappa & 0 & \theta_0\sigma^2_\kappa\\
0 & 0 & 0 & 0 & 0\\
0 & 0 & \theta_0\sigma^2_\kappa & 0 & \sigma^2_\xi + \theta^2_0\sigma^2_\kappa\\
\end{array}\right],
\label{eq:biv_og_model_general_ssf_mOmega}
\end{align}
with $\mathbf{c}=\mathbf{d}=0$. The transition matrix $\bT$ and the states variance--covariance matrix $\bOmega$ have been obtained by using the companion form of the AR(2) process and by substituting for the output gap measure $\psi_t$ in equation \eqref{eq:biv_og_model_specification_6} by its definition in equation \eqref{eq:biv_og_model_specification_4} obtaining the following relationship
\begin{equation}
\tau_t=\tau_{t-1}+\left(\theta_0\phi_1+\theta_1\right)\psi_{t-1}+\left(\theta_0\phi_2\right)\psi_{t-2}+\theta_0\kappa_t+\xi_t,
\label{eq:biv_og_model_inf_ssf_final}
\end{equation}
where $\theta_0\phi_1+\theta_1$ and $\theta_0\phi_2$ are the inflation loading factors and the innovation term $\theta_0\kappa_t+\xi_t$ is a linear combination of the cycle disturbance and the inflation first--difference dynamics. As in the original Kuttner's \citeyearpar{kuttner.1994} model the innovations in inflation dynamics (equation, \ref{eq:biv_og_model_inf_ssf_final}) and in the output gap (equation, \ref{eq:biv_og_model_specification_4}) are contemporaneously correlated, and the parameters enter non--linearly in the state space formulation. This is the reason why we propose the new adaptive--MCMC algorithm to make Bayesian inference for this model.\newline
\indent We complete the model formulation, we state the complete--data likelihood. The complete--data likelihood of the unobservable components $\left(\boldsymbol{\alpha}_t\right)_{t=1}^T$ and all parameters $\Xi$ can be factorized as follows:
\begin{eqnarray}
&&\mathcal{L}\left(\left(\boldsymbol{\alpha}_t\right)_{t=1}^T,\Xi\mid\mathbf{y}\right)\nonumber\\
&&\propto\prod_{t=1}^Tf\left(\by_{t}\mid\boldsymbol{\alpha}_t,\bSigma\right)
f\left(\balpha_1\right)
\prod_{t=1}^{T-1}f\left(\boldsymbol{\alpha_{t+1}}\mid\boldsymbol{\alpha_t},\bOmega\right)\nonumber\\
&&\propto\exp\left\{-\frac{1}{2\kappa}\boldsymbol{\alpha}_1^\trasp\boldsymbol{\alpha}_1\right\}
\exp\left\{-\frac{1}{2}\sum_{t=1}^{T-1}\left(\boldsymbol{\alpha}_{t+1}-\mathbf{d}-\bT\boldsymbol{\alpha}_t\right)^\trasp\bOmega^{-1}\left(\boldsymbol{\alpha}_{t+1}-\mathbf{d}-\bT\boldsymbol{\alpha}_t\right)\right\},\nonumber\\
&&\times\vert\bSigma\vert^{-\frac{dT}{2}}\vert\bOmega\vert^{-\frac{pT}{2}}
\exp\left\{-\frac{1}{2}\sum_{t=1}^T\left(\mathbf{y}_t-\mathbf{c}-\bZ_t\boldsymbol{\alpha}_t\right)^\trasp\bSigma^{-1}\left(\mathbf{y}_t-\mathbf{c}-\bZ_t\boldsymbol{\alpha}_t\right)\right\}
\label{eq:ssm_complete_like}
\end{eqnarray}
where the specification of the state space relevant matrices $\left(\bc,\mathbf{d},\bZ,\bT,\bSigma,\bOmega\right)$ is that defined above. 
%
\section{Prior specification}
\label{sec:prior_specification}
%
\noindent Before proceeding to the estimation of the models, we specify the prior assumptions on models parameters which represents an important ingredient of the Bayesian model specification procedure. Specifying a prior distribution for unobservable component models entails the choice of a family of distributions for each group of parameters, and the additional elicitation of the prior hyper parameters. When dealing with unobserved component models, it is important to recognise that to guarantee the posterior to be a proper distribution, priors should be proper (see e.g. Fr\"uhwirth--Schnatter, \citeyear{fruhwirth-schnatter.2006}). In addition, to prevent problems with the likelihood flatness and those related to model selection procedure, we avoid being fully non informative on all parameters. The specification of prior parameters may be particularly difficult when the parameter set is large and one of the main aims of the research is to compare different models in terms of their predictive ability, (see e.g. Geweke and Whiteman, \citeyear{geweke_whiteman.2006}), as in the case we consider here.\newline 
\indent Let us focus on the general bivariate model specified in equations \eqref{eq:biv_og_model_specification_1}--\eqref{eq:biv_og_model_specification_6} and let $\Xi=\left(\phi_1,\phi_2,\theta_0,\theta_1,\sigma^2_\eta,\sigma^2_\zeta,\sigma^2_\kappa,\sigma_\varepsilon^2,\sigma^2_\epsilon,\sigma^2_\xi\right)$ denote the vector of hyper parameters. We consider the following block conditional independent prior structure for the bivariate OG model encompassing all the considered specifications:
\begin{align}
\label{eq:prior_meas_rgdp}
\pi\left(\sigma_\epsilon^2\right)&\approx\invgamma\left(a_\epsilon,b_\epsilon\right)\\
\label{eq:prior_trend_rgdp}
\pi\left(\sigma_\eta^2\right)&\approx\invgamma\left(a_\eta,b_\eta\right)\\
\label{eq:prior_slope_rgdp}
\pi\left(\sigma_\zeta^2\right)&\approx\invgamma\left(a_\zeta,b_\zeta\right)\\
\label{eq:prior_cycle_var}
\pi\left(\sigma_\kappa^2\right)&\approx\invgamma\left(a_\kappa,b_\kappa\right)\\
\label{eq:prior_meas_infl}
\pi\left(\sigma_\varepsilon^2\right)&\approx\invgamma\left(a_\varepsilon,b_\varepsilon\right)\\
\label{eq:prior_trend_infl}
\pi\left(\sigma_\xi^2\right)&\approx\invgamma\left(a_\xi,b_\xi\right)\\
\label{eq:prior_rho}
\pi\left(\rho\right)&\approx\mathcal{B}e\left(a_\rho,b_\rho\right)\\
\label{eq:prior_lambda}
\pi\left(\frac{\lambda}{\pi}\right)&\approx\mathcal{B}e\left(a_\lambda,b_\lambda\right)\\
\label{eq:prior_theta_0}
\pi\left(\theta_0\right)&\approx\mathcal{N}\left(\mu_0,\sigma_0^2\right)\\
\label{eq:prior_theta_1}
\pi\left(\theta_1\right)&\approx\mathcal{N}\left(\mu_1,\sigma_1^2\right),
\end{align}
where $\mathcal{N}\left(\cdot,\cdot\right)$, $\mathcal{IG}\left(\cdot,\cdot\right)$ and $\mathcal{B}e\left(\cdot,\cdot\right)$ denotes the Normal, Inverse Gamma and Beta distributions, respectively. The priors for the hyper parameters $\left(\sigma_\epsilon^2,\sigma^2_\eta,\sigma^2_\zeta,\sigma^2_\varepsilon\right)$ have been chosen Inverse Gamma because they lead to naturally conjugate full conditional distributions. The conjugate property is instead lost for all the remaining hyper parameters mainly because of the fact that equation \eqref{eq:biv_og_model_specification_4} and equation\eqref{eq:biv_og_model_specification_6} are correlated. For the remaining scale parameters the Inverse Gamma and the Gaussian distributions are the natural candidates for the scales $\left(\sigma_\xi^2,\sigma_\kappa^2\right)$ and the loadings parameters $\left(\theta_0,\theta_1\right)$, respectively. 
According to Planas \textit{et al.} \citeyearpar{planas_etal.2008}, Harvey \textit{et al.} \citeyearpar{harvey_etal.2007} and Proietti \citeyearpar{proietti.2009} we choose a Beta prior for the cycle amplitude $\rho$ and the frequency parameter $\lambda$ rescaled to belong to the set of elicitable frequencies $\left(0,\pi\right)$. This prior structure allows enough flexibility in the signal estimation and extraction while preserving a proper posterior distribution. The amount of prior information used to estimate parameters and latent processes can be tuned by appropriately choosing the prior parameters $\left(a_\epsilon,b_\epsilon,a_\eta,b_\eta,a_\zeta,b_\zeta,a_\kappa,b_\kappa,a_\varepsilon,b_\varepsilon,a_\xi,b_\xi,a_\rho,b_\rho,a_\lambda,b_\lambda,\mu_0,\sigma_0^2,\mu_1,\sigma_1^2\right)$. 
%
\begin{table}
\centering
\captionsetup{font={footnotesize}, labelfont=sc}  
\tabcolsep=2.0mm
\begin{small}
\begin{tabular}{cccccl} 
\toprule
\multirow{2}{*}{Parameter} & \multicolumn{2}{c}{Hyperparameters} & \multicolumn{2}{c}{Prior Moments} & \multirow{2}{*}{Description}\\
\cmidrule(lr){2-3}\cmidrule(lr){4-5}
& Location & Scale & Mean & Std. Dev.     &    \\
\cmidrule(lr){1-1}\cmidrule(lr){2-3}\cmidrule(lr){4-5}\cmidrule(lr){6-6}
$\sigma_{\epsilon}^2$ 	& 3.0		& $3.6\times10^{-3}$		& $1.8\times10^{-3}$& $1.8\times10^{-3}$	  &  $y_t$ innovation variance       \\
$\sigma_{\eta}^2$ 		& 3.0		& $4.0\times10^{-3}$		&  $2.0\times10^{-3}$	&      $2.0\times10^{-3}$     	& $y_t$ trend variance     \\
$\sigma_{\zeta}^2$ 		& 3.0		& $4.0\times10^{-4}$		& $2.0\times10^{-4}$	&          $2.0\times10^{-4}$	& $y_t$ slope variance     \\
$\sigma_{\varepsilon}^2$ 	& 3.0		& $3.2\times10^{-4}$		& $1.6\times10^{-4}$ 	& $1.6\times10^{-4}$  &  $\Delta p_t$ innovation variance       \\
$\sigma_{\xi}^2$ 		& 3.0 	& $3.2\times10^{-4}$ 		&  $1.6\times10^{-4}$	&      $1.6\times10^{-4}$      & $\Delta p_t$ trend variance     \\
$\sigma_{\kappa}^2$ 	& 2.40		& 2.82		& 2.01	&   10.0       	&  cycle variance     \\
$\rho$ 				& 2.0		& 3.0 		& 0.4		& 0.04          	&  cycle amplitude     \\
$\lambda$ 			& 5.54		& 27.72	& $\frac{2\pi}{12}$	& $0.20^{2}$		&  cycle frequency     \\
$\theta_0$ 			& 0.0		& 100.0 	& 0.0		& 10.0         	&  $\Delta p_t$ OG loading      \\
$\theta_1$ 			& 0.0 	& 100.0 	& 0.0		& 10.0        	&  $\Delta p_t$ OG loading     \\
\bottomrule
\end{tabular}
\end{small}
\caption{\footnotesize{Prior hyper parameters for the general bivariate OG model defined in equations \eqref{eq:biv_og_model_specification_1}--\eqref{eq:biv_og_model_specification_6}.}}
\label{tab:prior_hyperparameters}
\end{table}
%
The previous prior structure can be adapted to the different model specifications considered in this paper. For example, in the case of univariate local level model plus cycle only the priors for the parameters involved in that specification, $\left(\sigma_\epsilon^2,\sigma_\eta^2,\sigma_\kappa^2,\rho,\lambda\right)$ are retained form the above prior structure. Concerning the prior hyper parameter elicitation, we consider all the available information form macroeconomic theory, other series and previous works on business cycle analysis. Table \ref{tab:prior_hyperparameters} summarises for each parameter the corresponding hyper parameters (location and scale) and the implied prior means and standard deviations. Regarding the cycle amplitude $\rho$ and frequency $\lambda$, which are the most critical parameters, we take into account the results of Zizza \citeyearpar{zizza.2006} that identifies for the Italian OG a period ranging from 3 to 5 years and an amplitude from $0.12$ to $0.30$ (as a percentage of the trend GDP), using structural time series models. Gallegati and Stanca \citeyearpar{gallegati_stanca.1998} found similar result using the HP filter for about the same period. We, thus, centre the first moment of the $\rho$ and $\lambda$ prior distributions on these values without imposing too much precision. Specifically, we set the mean and the standard deviation of the cycle frequency $\lambda$ equal to $\frac{2\pi}{12}$ and $0.20$, respectively. This corresponds to a mean period for the business cycle of about 3 years. The $95\%$ high prior density credible interval goes from about 2 years to more than 9 years, covering the most probable period lengths without imposing too much prior information on the business cycle duration. Moreover, the support of the cycle frequency is extended to non--elicitable frequencies. The prior amplitude mean and standard deviation are set equal to $0.4$ and $0.04$ respectively, with a $95\%$ high prior density credible interval going from about $0.04$ to $0.78$. Concerning the Phillips curve equation, the prior distribution of the OG loading parameters $\left(\theta_0,\theta_1\right)$ can, in principle, be tailored on the basis of the available information from both theoretical and empirical works. For example, the inflation is expected to react positively to changes in the output gap and the contemporaneous correlation between shocks in the output gap and shocks in inflation, measured by the parameter $\theta_0$ should also be positive. However, as argued in the introduction, since one of the main objectives of this paper is to test empirically the validity of the Phillips curve we choose a sufficiently diffuse prior for the OG loadings centred around zero.\newline
%
%
\indent Concerning the specification of the scale innovations and trends hyperparameters, since any information is available from previous works on the italian output gap, we decided to be as non--informative as possible. In particular, as shown in Table \ref{tab:prior_hyperparameters} the real GDP innovation variance $\sigma_\epsilon^2$ is larger on average than the inflation variance to account for the larger variance the series displays during the first part of the sample. Furthermore, real GDP variance has a priori standard deviation which is ten time larger than that of the inflation variance. To elicitate the trend and slope prior variance instead we adopt the strategy. As indicated in Table \ref{tab:prior_hyperparameters}, we adopt prior mean and standard deviation for the variance of the real GDP trend $\sigma_\eta^2$ ten time larger than those imposed on the inflation trend $\sigma^2_\xi$.  
%
%
%
%
\section{Bayesian inference}
\label{sec:og_inference}
%
During the past decades Markov Chain Monte Carlo (MCMC) methods, Metropolis \textit{et al.} \citeyearpar{metropolis_etal.1953} and Hastings \citeyearpar{hastings.1970}, have been extensively developed within the Bayesian approach to sample from analytically intractable posterior distributions with particular emphasis to the Gibbs Sampler and the Metropolis--Hastings algorithms. 
In the context of state space models, Bayesian methods have been introduced by West and Harrison \citeyearpar{west_harrison.1997} and subsequently considered by Carter and Kohn (\citeyear{carter_kohn.1994}, \citeyear{carter_kohn.1996}) and Fr\"{u}hwirth-Schnatter \citeyearpar{fruhwirth-schnatter.1994}, Durbin and Koopman \citeyearpar{durbin_koopman.2002} to develop multi--move sampler to jointly simulate the entire set of latent states conditional on model parameters. 
%
%
%
Durbin and Koopman \citeyearpar{durbin_koopman.2000} recently provided a comprehensive treatment of estimation, filtering and smoothing of state space models from both a classic and Bayesian perspective. If the model includes fixed unknown parameters, a standard technique to perform parameter estimation consists of augmenting the posterior distribution of the parameters with the joint latent states, and then marginalising out numerically the latent variables using simulation techniques. This data augmentation technique relies on the availability of optimal filters to simulate from the joint full conditional distribution of the latent states conditional on the parameters, such as the Forward--Filtering--Backward--Smoothing (FFBS) algorithm of Carter and Kohn (\citeyear{carter_kohn.1994}, \citeyear{carter_kohn.1996}) and Fr\"{u}hwirth-Schnatter \citeyearpar{fruhwirth-schnatter.1994}. See, for example Fearnhead \citeyearpar{fearnhead.2011} and Robert and Casella \citeyearpar{robert_casella.2004} and for detailed introductions to MCMC techniques that apply to state space models. However, this approach usually requires to revert to MCMC techniques to simulate parameters conditional to the states previously drawn, which suffer from several drawbacks. The full conditional distribution of the parameters are often intractable and to simulate directly from them is impossible. This essentially means that we should revert to Metropolis--within--Gibbs methods implying longer computational time and an increased probability that the simulator remains confined within local modes and fails to explore the entire support of the joint posterior. Furthermore, Metropolis--type algorithms requires the scale parameters to be calibrated in advance to get acceptable acceptance rates. The calibration of M.--H. algorithms is quite complicated because it is usually done by simulating samples of small size with different scales and then comparing the resulting acceptance rates of chains which would not converged yet. Those acceptance rates are often different form those we usually get on initial draws and this makes the process most of the time useless for the purpose of calibrating the proposal scale. Moreover, due to the usually large dimension of the parameter space and the multimodality of the posterior distributions arising in this context, standard MCMC algorithms such that the Metropolis--Hastings \citeyearpar{hastings.1970} and the Gibbs sampler of Geman and Geman \citeyearpar{geman_geman.1984}, usually fail to explore the entire support of the posterior distribution.
For a deep and up to date discussion on MCMC methods for general state space methods we refer to the books Capp\'e \textit{et al.} \citeyearpar{cappe_etal.2005} and Fr\"uwirth-Schnatter \citeyearpar{fruhwirth-schnatter.2006}, while Bayesian optimal filtering and smoothing techniques are extensively analysed by S\"arkk\"a \citeyearpar{sarkka.2013} and Challa \textit{et al.} \citeyearpar{challa_etal.2011}.\newline
\indent Here we follow a different adaptive approach developed in Bernardi (\citeyear{bernardi.2011a}, \citeyear{bernardi.2011b}) to simulate from the posterior distribution of finite mixture and stochastic volatility models. 
Adaptive MCMC methods extend traditional Metropolis--Hastings sampling schemes by allowing the proposal distribution to change over time in such a way that previous draws from the chain are used to tailored the proposal distributions to minimise the distance with the target. For an up to date introduction on Adaptive--MCMC, see e.g. Andrieu and Thoms \citeyearpar{andrieu_thoms.2008} and Liang \textit{et al.} \citeyearpar{liang_etal.2010}.
Basically, Adaptive--MCMC schemes constructs a Markov chain whose equilibrium distribution asymptotically converges to the joint posterior of the model parameters, here $\pi\left(\boldsymbol{\Xi}\mid\by\right)$. After running the Markov chain for a burn--in period, one obtains samples from the limiting distribution, as in the case of standard MCMC methods, provided that the Markov chain has reached convergence, see e.g. Atchad\'e and Rosenthal \citeyearpar{atchade_rosenthal.2005}
Atchad\'e \textit{et al.} \citeyearpar{atchade_etal.2011}. The convergence and ergodicity of these algorithms is guaranteed by a careful design of the adaptation strategy, see Roberts and Rosenthal \citeyearpar{roberts_rosenthal.2007}.\newline
\indent In this paper we propose an adaptive Metropolis--within--Gibbs sampler where the parameter space is augmented by the latent states $\balpha=\left\{\boldsymbol{\alpha}_t,t=1,2,\ldots,T\right\}^{(i)}$, where $\balpha_t=\left(\mu_t,\beta_t,\psi_t,\psi_{t-1},\tau_t\right)$ denotes the vector of unobservable states at time $t$, to get the joint posterior distribution of parameters and states $\pi\left(\boldsymbol{\Xi},\boldsymbol{\alpha}\mid\by\right)$. Latent states are subsequently marginalised out using the Gibbs sampler algorithm which consists of two main steps where we simulate alternatively from the full conditional distribution of the states given parameters and observations, $\pi\left(\boldsymbol{\alpha}\mid\boldsymbol{\Xi},\by\right)$, and from that of the parameters given the states $\pi\left(\boldsymbol{\Xi}\mid\boldsymbol{\alpha},\by\right)$. We draw the latent states $\balpha_t$ for $t=1,2,\dots,T$ jointly from the full conditional distribution $\pi\left(\boldsymbol{\alpha}\mid\by,\boldsymbol{\Xi}\right)$ using the multi--move simulation smoother of Durbin and Koopman (\citeyear{durbin_koopman.2002}, \citeyear{durbin_koopman.2012}). This entails running the Kalman filter forward for the state space structure defined in equations \eqref{eq:biv_og_model_general_ssf_mZ}--\eqref{eq:biv_og_model_general_ssf_mOmega}. Once the Kalman filter is run forward, we run the Kalman smoother backward in order to get the moments of joint full conditional distribution of the latent states (\ref{eq:ssm_complete_like}). Finally, we simulate a sample path by drawing from this joint distribution. For a similar simulation method based on forward--filtering backward--smoothing (FFBS) algorithm, see also Carter and Kohn (\citeyear{carter_kohn.1994}, \citeyear{carter_kohn.1996}) and Fr\"{u}hwirth-Schnatter \citeyearpar{fruhwirth-schnatter.1994}. The adaptation strategy is then applied to simulate from those parameters for which the full conditional distribution is not known. For each parameter or block of parameters a specific proposal distribution is chosen whose parameters are adapted to minimise Kullback--Leibler (KL) divergence from the target distribution. The method makes use of the stochastic approximation
approach of Robbins and Monro \citeyearpar{robbins_monro.1951} and is a generalisation of the Algoritm 2 presented in Andrieu and Thoms \citeyearpar{andrieu_thoms.2008}. For a detailed introduction on how to use stochastic approximations techniques to minimise the KL divergence see \citeyearpar{ji.2006} and Andrieu, Moulines \citeyearpar{andrieu_moulines.2006} and Bernardi \citeyearpar{bernardi.2011a}.
In the next subsection we details the Adaptive--MCMC sampler.
\subsection{The Adaptive Metropolis--within--Gibbs sampler}
\label{sec:gibbs_sampler}
%
\indent Let us focus on the general bivariate model defined in equations \eqref{eq:biv_og_model_specification_1}--\eqref{eq:biv_og_model_specification_6} and denote by $\by$ the stack of observations $\left\{y_t,\Delta p_t\right\}_{t=1}^T$, $\balpha=\left\{\boldsymbol{\alpha}_t,t=1,2,\ldots,T\right\}$, where $\balpha_t=\left(\mu_t,\beta_t,\psi_t,\psi_{t-1},\tau_t\right)$ denotes the vector of unobservable states at time $t$ and $\boldsymbol{\Xi}=\left(\phi_1,\phi_2,\theta_0,\theta_1,\sigma^2_\eta,\sigma^2_\zeta,\sigma^2_\kappa,\sigma_\varepsilon^2,\sigma^2_\epsilon,\sigma^2_\xi\right)$ denote the vector of hyper parameters. The approximation of the joint posterior density of the model parameters $\boldsymbol{\Xi}$ and states $\balpha$ is obtained by carrying out the following Adaptive--Independent Metropolis--Hastings--within--Gibbs (AIMH--GS, henceforth) algorithm. The AIMH--GS algorithm consists of the following steps.\newline\newline
\noindent After choosing a set of initial values for the parameter vector $\boldsymbol{\Xi}^{(0)}$, simulations from the posterior distribution at the $i$--th iteration of $\boldsymbol{\Xi}^{(i)}$, $\balpha^{(i)}=\left\{\boldsymbol{\alpha}_t,t=1,2,\ldots,T\right\}^{(i)}$, for $i=1,2,\dots$, are obtained by running iteratively the following steps: 
%
\begin{enumerate}
\item Generate $\sigma_\epsilon^2$ from $\mathcal{IG}\left(a_\epsilon^T,b^T_\epsilon\right)$ with parameters
%
\begin{equation} 
a_\epsilon^T=a_\epsilon+\frac{T}{2},\qquad b_\epsilon^T=b_\epsilon+\frac{1}{2}\sum_{t=1}^{T}\left(y_t-\mu_t^{(i)}-\psi_t^{(i)}\right)^2.
\end{equation}
%
\item Generate $\sigma_\varepsilon^2$ from $\mathcal{IG}\left(a_\epsilon^T,b^T_\epsilon\right)$ with parameters
\begin{equation} 
a_\epsilon^T=a_\epsilon+\frac{T}{2},\qquad b_\epsilon^T=b_\epsilon+\frac{1}{2}\sum_{t=1}^{T}\left(\pi_{t}-\tau_t^{(i)}\right)^2.
\end{equation}
%
\item Generate $\sigma_\eta^2$ from $\mathcal{IG}\left(a_\eta^T,b^T_\eta\right)$ with parameters
\begin{equation} 
a_\eta^T=a_\eta+\frac{T-1}{2},\qquad b_\eta^T=b_\eta+\frac{1}{2}\sum_{t=1}^{T-1}\left(\mu_{t+1}^{(i)}-\mu_{t}^{(i)}-\beta_{t}^{(i)}\right)^2.
\end{equation}
%
\item Generate $\sigma_\zeta^2$ from $\mathcal{IG}\left(a_\zeta^T,b^T_\zeta\right)$ with parameters
\begin{equation} 
a_\zeta^T=a_\zeta+\frac{T-1}{2},\qquad b_\zeta^T=b_\zeta+\frac{1}{2}\sum_{t=1}^{T-1}\left(\beta_{t+1}^{(i)}-\beta_{t}^{(i)}\right)^2.
\end{equation}
%
\item Generate $\theta_1^*$ from the following independent Metropolis--Hastings proposal distribution $p^{(i)}\left(\theta_1^*,\theta_1^c\right)\approx\mathcal{N}\left(\theta_1\mid\mu_1^{(i)},\sigma_1^{(i)}\right)$ at iteration $i$, and accept the candidate $\theta_1^*$ with the following acceptance probability
\begin{eqnarray}
\gamma^{(i)}_{\theta_1}
\left(\theta_1^*,\theta_1^{c}\right)=\min\left[\frac{\pi_\tau\left(\btau\mid\by,\theta_1^*\right)
\pi\left(\theta_1^*\right)
p^{(i)}\left(\theta_1^*,\theta_1^c\right)}{\pi_\tau\left(\btau\mid\by,\theta_1^c\right)\pi\left(\theta_1^c\right)p^{(i)}\left(\theta_1^c,\theta_1^*\right)},1\right],
\end{eqnarray}
where $\pi_\tau\left(\btau^{(i)}\mid\by,\theta_1\right)$ is the posterior distribution of the inflation trend $\btau=\left(\tau_1,\tau_2,\dots,\tau_{T+1}\right)$ and $\pi\left(\theta_1\right)$ is the prior distribution of the parameter $\theta_1$ defined in the previous section. The distribution of the trend--inflation latent states $\tau$'s is the only one which involves the loading factor $\theta_1$. The full conditional posterior distribution $\pi_\tau\left(\btau^{(i)}\mid\by,\theta_1\right)$ can be easily calculated from the joint distribution in equation \eqref{eq:ssm_complete_like}, and has the following form
\begin{eqnarray}
\pi_\tau\left(\btau\mid\by,\theta_1\right)=\frac{\exp\left\{-\frac{1}{2}\sum_{t=1}^{T-1}\left[\nu_{t+1}^{(i)}\left(\theta_1\right)\right]^2\right\}}{\left(\sigma_\xi^2+\theta_0^2\sigma_\kappa^2\right)^{\frac{T-1}{2}}},
\label{eq:full_conditional_posterior_theta_1}
\end{eqnarray}
where $\nu_{t+1}^{(i)}\left(\theta_1\right)=\Delta\tau_{t+1}^{(i)}-\left(\theta_0^{(i)}\phi_1^{(i)}+\theta_1\right)\psi_t^{(i)}-\theta_0^{(i)}\phi_2^{(i)}\psi_{t-1}^{(i)}$ and $\Delta\tau_{t+1}=\tau_{t+1}-\tau_t$.
After the simulation step takes place, we adapt the proposal parameters using the following equations:
\begin{eqnarray}
\mu^{(i+1)}_1&=&\mu^{(i)}_1+\delta^{(i+1)}\left(\theta_1-\mu^{(i)}_1\right)
\label{eq:theta_1_adapt_mean}\\
\sigma^{(i+1)}_1&=&\sigma^{(i)}_1+\delta^{(i+1)}\left(\theta_1-\mu^{(i+1)}_1\right)^2,
\label{eq:theta_1_adapt_var}
\end{eqnarray}
as in Andrieu and Thoms \citeyearpar{andrieu_thoms.2008}, where hereafter $\delta^{(i+1)}$ denotes a tuning parameter that should be carefully selected to ensure the convergence and ergodicity of the resulting chain, see Andrieu and Moulines \citeyearpar{andrieu_moulines.2006} and the discussion at the end of the algorithm.
%
\item Generate $\theta_0^*$ from the following independent Metropolis--Hastings proposal distribution $p^{(i)}\left(\theta_0^*,\theta_0^c\right)\approx\mathcal{N}\left(\theta_1\mid\mu_0^{(i)},\sigma_0^{(i)}\right)$ at iteration $i$, and accept the candidate $\theta_0^*$ with the following acceptance probability
\begin{eqnarray}
\gamma^{(i)}_{\theta_0}
\left(\theta_0^*,\theta_0^{c}\right)=\min\left[\frac{\pi_{\tau,\psi}\left(\btau,\bpsi\mid\by,\theta_0^*\right)
\pi\left(\theta_0^*\right)
p^{(i)}\left(\theta_0^*,\theta_0^c\right)}{\pi_{\tau,\psi}\left(\btau,\bpsi\mid\by,\theta_0^c\right)\pi\left(\theta_0^c\right)p^{(i)}\left(\theta_0^c,\theta_0^*\right)},1\right],
\end{eqnarray}
where $\pi_{\tau,\psi}\left(\btau^{(i)},\bpsi^{(i)}\mid\by,\theta_0\right)$ is the posterior distribution of the inflation trend $\btau=\left(\tau_1,\tau_2,\dots,\tau_{T+1}\right)$ and the cycle $\bpsi=\left(\psi_1,\psi_2,\dots,\psi_{T+1}\right)$ and $\pi\left(\theta_0\right)$ is the prior distribution of the parameter $\theta_1$ defined in the previous section. The full conditional posterior distribution $\pi_\tau\left(\btau^{(i)}\mid\by,\theta_1\right)$ can be easily calculated from the joint distribution in equation \eqref{eq:ssm_complete_like}, and has the following form
{\small
\begin{eqnarray}
\pi_{\tau,\psi}\left(\btau,\bpsi\mid\by,\theta_1\right)=\frac{
\exp\left\{-\frac{1}{2}\sum_{t=1}^{T-1}\left[\bnu_{t+1}^{(i)}\left(\theta_0\right)\right]^\trasp\Sigma\left(\theta_0\right)^{-1}
\left[\bnu_{t+1}^{(i)}\left(\theta_0\right)\right]\right\}}{\vert\Sigma\left(\theta_0\right)\vert^{\frac{T-1}{2}}},
\label{eq:full_conditional_posterior_theta_0}
\end{eqnarray}}
where 
\begin{eqnarray}
\bnu_{t+1}^{(i)}\left(\theta_0\right)&=&
\left[\begin{array}{c}
\psi_{t+1}^{(i)}-\phi_1^{(i)}\psi_{t}^{(i)}-\phi_2^{(i)}\psi_{t-1}^{(i)}\\
\Delta\tau_{t+1}^{(i)}-\left(\theta_0^{(i)}\phi_1^{(i)}+\theta_1\right)\psi_t^{(i)}-\theta_0^{(i)}\phi_2^{(i)}\psi_{t-1}^{(i)}
\end{array}
\right]\\
\Sigma\left(\theta_0\right)&=&\left[
\begin{array}{cc}
\sigma_\kappa^2 & \theta_0\sigma_\kappa^2\\
\theta_0\sigma_\kappa^2 & \theta_0^2\sigma_\kappa^2+\sigma_\xi^2
\end{array}
\right].
\end{eqnarray}
After the simulation step takes place, we adapt the proposal parameters using previous equations (\ref{eq:theta_1_adapt_mean}--\ref{eq:theta_1_adapt_mean}).
%
\item Generate $\sigma_\xi^{*}$ from the following independent Metropolis--Hastings proposal distribution $p^{(i)}\left(\sigma_\xi^{2,*},\sigma_\xi^{2,c}\right)\approx\mathcal{IG}\left(a_\xi,b_\xi\right)$ at iteration $i$, and accept the candidate $\sigma_\xi^{2,*}$ with the following acceptance probability
\begin{eqnarray}
\gamma^{(i)}_{\sigma_\xi^2}
\left(\sigma_\xi^{2,*},\sigma_\xi^{2,c}\right)=\min\left[\frac{\pi_\tau\left(\btau\mid\by,\sigma_\xi^{2,*}\right)
\pi\left(\sigma_\xi^{2,*}\right)
p^{(i)}\left(\sigma_\xi^{2,*},\sigma_\xi^{2,c}\right)
}{\pi_\tau\left(\btau\mid\by,\sigma_\xi^{2,*}\right)
\pi\left(\sigma_\xi^{2,*}\right)
p^{(i)}\left(\sigma_\xi^{2,*},\sigma_\xi^{2,c}\right)},1\right],
\end{eqnarray}
where $\pi_\tau\left(\btau^{(i)}\mid\by,\sigma_\xi^2\right)$ is the full conditional posterior distribution of the inflation trend $\btau=\left(\tau_1,\tau_2,\dots,\tau_{T+1}\right)$ defined in the previous equation \eqref{eq:full_conditional_posterior_theta_1} and $\pi\left(\sigma_\xi^2\right)$ is the prior distribution of the parameter $\sigma_\xi^2$ defined in the previous section. The proposal distribution parameters are then adapted through the following equations:
\begin{eqnarray}
a_\xi^{(i+1)}&=&a_\xi^{(i)}+\delta^{(i+1)}\left[\log\left(\frac{b_\xi^{(i)}}{\sigma_\xi^2}\right)-\Psi\left(\sigma_\xi^2\right)\right]\nonumber\\
b_\xi^{(i+1)}&=&b_\xi^{(i)}+\delta^{(i+1)}\left[\frac{a_\xi^{(i)}}{b_\xi^{(i)}}-\frac{1}{\sigma_\xi^2}\right],\nonumber
\end{eqnarray}
where $\Psi\left(\cdot\right)$ is the digamma function.
%
\item Generate $\rho^{*}$ from the following independent Metropolis--Hastings proposal distribution $p^{(i)}\left(\rho^*,\rho^{c}\right)\approx\mathcal{B}\left(a_\rho,b_\rho\right)$ at iteration $i$, and accept the candidate $\rho^{*}$ with the following acceptance probability
\begin{eqnarray}
\gamma^{(i)}_{\rho}
\left(\rho^*,\rho^{c}\right)=\min\left[\frac{\pi_{\tau,\psi}\left(\btau,\bpsi\mid\by,\rho^*\right)
\pi\left(\rho^*\right)
p^{(i)}\left(\rho^*,\rho^c\right)}{\pi_{\tau,\psi}\left(\btau,\bpsi\mid\by,\rho^c\right)\pi\left(\rho^c\right)p^{(i)}\left(\rho^c,\rho^*\right)},1\right],
\end{eqnarray}
where $\pi_{\tau,\psi}\left(\btau^{(i)},\bpsi^{(i)}\mid\by,\rho\right)$ is the posterior distribution of the inflation trend $\btau=\left(\tau_1,\tau_2,\dots,\tau_{T+1}\right)$ and the cycle $\bpsi=\left(\psi_1,\psi_2,\dots,\psi_{T+1}\right)$ defined in equation \eqref{eq:full_conditional_posterior_theta_0} and $\pi\left(\rho\right)$ is the prior distribution of the parameter $\rho$ defined in the previous section. The updating equations becomes:
\begin{eqnarray}
a_\rho^{(i+1)}&=&a_\rho^{(i)}+\delta^{(i+1)}\left[\log\left(\rho\right)+\Psi\left(a_\rho^{(i)}+b_\rho^{(i)}\right)-\Psi\left(a_\rho^{(i)}\right)\right]\nonumber\\
b_\rho^{(i+1)}&=&b_\rho^{(i)}+\delta^{(i+1)}\left[\log\left(1-\rho\right)+\Psi\left(a_\rho^{(i+1)}+b_\rho^{(i)}\right)-\Psi\left(b_\rho^{(i)}\right)\right].\nonumber
\end{eqnarray}
%
\item Generate $\lambda^{*}$ from the following independent Metropolis--Hastings proposal distribution $p^{(i)}\left(\lambda^*,\lambda^{c}\right)\approx\mathcal{B}\left(a_\lambda,b_\lambda\right)\bbone_{\left(0,2\pi\right)}\left(\lambda\right)$ at iteration $i$, and accept the candidate $\lambda^{*}$ with the following acceptance probability
\begin{eqnarray}
\gamma^{(i)}_{\lambda}
\left(\lambda^*,\lambda^{c}\right)=\min\left[\frac{\pi_{\tau,\psi}\left(\btau,\bpsi\mid\by,\lambda^*\right)
\pi\left(\lambda^*\right)
p^{(i)}\left(\lambda^*,\lambda^c\right)}{\pi_{\tau,\psi}\left(\btau,\bpsi\mid\by,\lambda^c\right)\pi\left(\lambda^c\right)p^{(i)}\left(\lambda^c,\lambda^*\right)},1\right],
\end{eqnarray}
where $\pi_{\tau,\psi}\left(\btau^{(i)},\bpsi^{(i)}\mid\by,\lambda\right)$ is the posterior distribution of the inflation trend $\btau=\left(\tau_1,\tau_2,\dots,\tau_{T+1}\right)$ and the cycle $\bpsi=\left(\psi_1,\psi_2,\dots,\psi_{T+1}\right)$ defined in equation \eqref{eq:full_conditional_posterior_theta_0} and $\pi\left(\lambda\right)$ is the prior distribution of the parameter $\lambda$ defined in the previous section. The updating equations becomes:
\begin{eqnarray}
a_\lambda^{(i+1)}&=&a_\lambda^{(i)}+\delta^{(i+1)}\left[\log\left(\frac{\lambda}{2\pi}\right)+\Psi\left(a_\lambda^{(i)}+b_\lambda^{(i)}\right)-\Psi\left(a_\lambda^{(i)}\right)\right]\nonumber\\
b_\lambda^{(i+1)}&=&b_\lambda^{(i)}+\delta^{(i+1)}\left[\log\left(\frac{2\pi-\lambda}{2\pi}\right)+\Psi\left(a_\lambda^{(i+1)}+b_\lambda^{(i)}\right)-\Psi\left(b_\lambda^{(i)}\right)\right].\nonumber
\end{eqnarray}
%
\item Generate $\sigma_\kappa^{*}$ from the following independent Metropolis--Hastings proposal distribution $p^{(i)}\left(\sigma_\kappa^{2,*},\sigma_\kappa^{2,c}\right)\approx\mathcal{IG}\left(a_\kappa,b_\kappa\right)$ at iteration $i$, and accept the candidate $\sigma_\kappa^{2,*}$ with the following acceptance probability
\begin{eqnarray}
\gamma^{(i)}_{\sigma_\kappa^2}
\left(\sigma_\kappa^{2,*},\sigma_\kappa^{2,c}\right)=\min\left[\frac{\pi_{\tau,\psi}\left(\btau,\bpsi\mid\by,\sigma_\kappa^{2,*}\right)
\pi\left(\sigma_\kappa^{2,*}\right)
p^{(i)}\left(\sigma_\kappa^{2,*},\sigma_\kappa^{2,c}\right)}{\pi_{\tau,\psi}\left(\btau,\bpsi\mid\by,\sigma_\kappa^{2,c}\right)\pi\left(\sigma_\kappa^{2,c}\right)p^{(i)}\left(\sigma_\kappa^{2,c},\sigma_\kappa^{2,*}\right)},1\right],
\end{eqnarray}
where $\pi_{\tau,\psi}\left(\btau,\bpsi\mid\by,\sigma_\kappa^{2,*}\right)$  is the posterior distribution of the inflation trend $\btau=\left(\tau_1,\tau_2,\dots,\tau_{T+1}\right)$ and the cycle $\bpsi=\left(\psi_1,\psi_2,\dots,\psi_{T+1}\right)$ defined in equation \eqref{eq:full_conditional_posterior_theta_0} and $\pi\left(\sigma_\kappa^2\right)$ is the prior distribution of the parameter $\sigma_\kappa^2$ defined in the previous section. The proposal distribution parameters are then adapted through the following equations:
\begin{eqnarray}
a_\kappa^{(i+1)}&=&a_\kappa^{(i)}+\delta^{(i+1)}\left[\log\left(\frac{b_\kappa^{(i)}}{\sigma_\kappa^2}\right)-\Psi\left(\sigma_\kappa^2\right)\right]\nonumber\\
b_\kappa^{(i+1)}&=&b_\kappa^{(i)}+\delta^{(i+1)}\left[\frac{a_\kappa^{(i)}}{b_\kappa^{(i)}}-\frac{1}{\sigma_\kappa^2}\right],\nonumber
\end{eqnarray}
where $\Psi\left(\cdot\right)$ is the digamma function.
%
\item As previously discussed, we simulate the latent states $\balpha_t^{(i)}$ for $t=1,2,\dots,T$ jointly from the full conditional distribution $\pi\left(\boldsymbol{\alpha}\mid\by,\boldsymbol{\Xi}\right)$ using the multi--move simulation smoother of Durbin and Koopman (\citeyear{durbin_koopman.2002}, \citeyear{durbin_koopman.2012}).
\end{enumerate}
At each iteration of the above sampler, the parameter $\delta$ should be tuned in order ensure the convergence and ergodicity property with respect the joint posterior distribution of the resulting chain. 
Roberts and Rosenthal \citeyearpar{roberts_rosenthal.2007} provide two conditions for the convergence of the chain: the diminishing adaptation condition, which is satisfied if and only if $\delta^{(i)}\longrightarrow 0$, as $i\rightarrow+\infty$, and the bounded convergence condition, which essentially guarantees that all transition kernels considered have bounded convergence time. Andrieu and Moulines \citeyearpar{andrieu_moulines.2006} show that both conditions are satisfied if and only if $\delta^{(i)}\propto i^{-d}$ where $d\in\left[0.5,1\right]$. We choose $\delta=\frac{1}{Ci^{0.5}}$ where $C$ is set to 10, i.e. $C=10$. As argued by Roberts and Rosenthal \citeyearpar{roberts_rosenthal.2007}, together these two conditions ensure asymptotic convergence and a
weak law of large numbers for this algorithm.
%
%
\section{Extracting the Italian output gap}
\label{sec:application}
%
In this section we give details about the data used to estimate the models described in the previous Section \ref{sec:og_models} for the Italian trend--cycle decomposition and we describe the main results about the output of the inference on the parameters of the different model specification we consider in this work.
%
\begin{figure}[!t]
\begin{center}
\captionsetup{font={small}, labelfont=sc}
\includegraphics[width=1.0\linewidth]{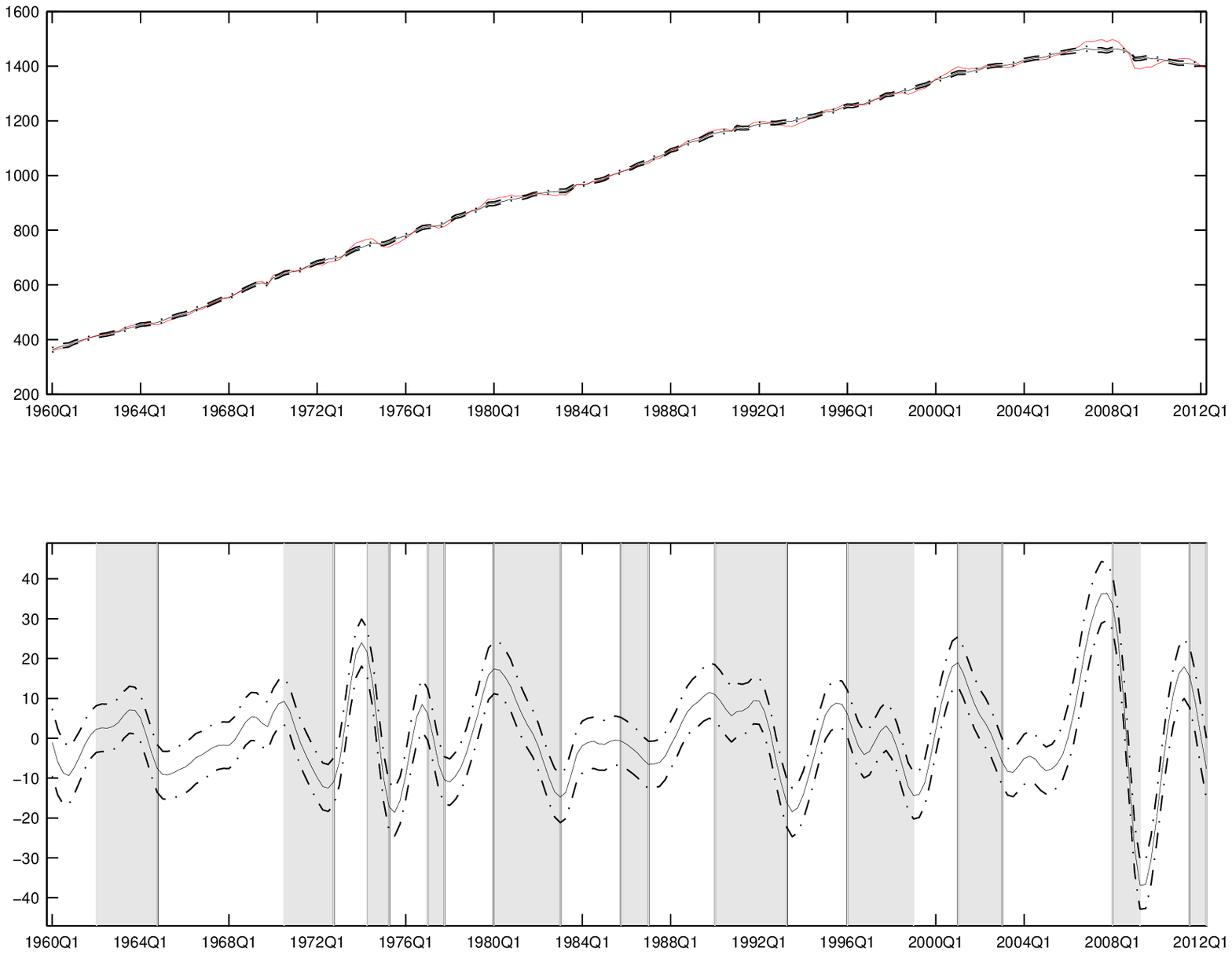}
\caption{\footnotesize{Posterior trend \textit{(upper panel)} and cycle \textit{(bottom panel)} estimates for the model $\mathcal{M}_{U}^{\sL\sL}$. The shaded area represents $95\%$ credible sets while the red line in the upper figure represents the real GDP.}}
\label{fig:univ_OG_model_LLM_AR2_trend_cycle}
\end{center}
\end{figure}

\begin{figure}[!ht]
\begin{center}
\captionsetup{font={small}, labelfont=sc}
\includegraphics[width=1.0\linewidth]{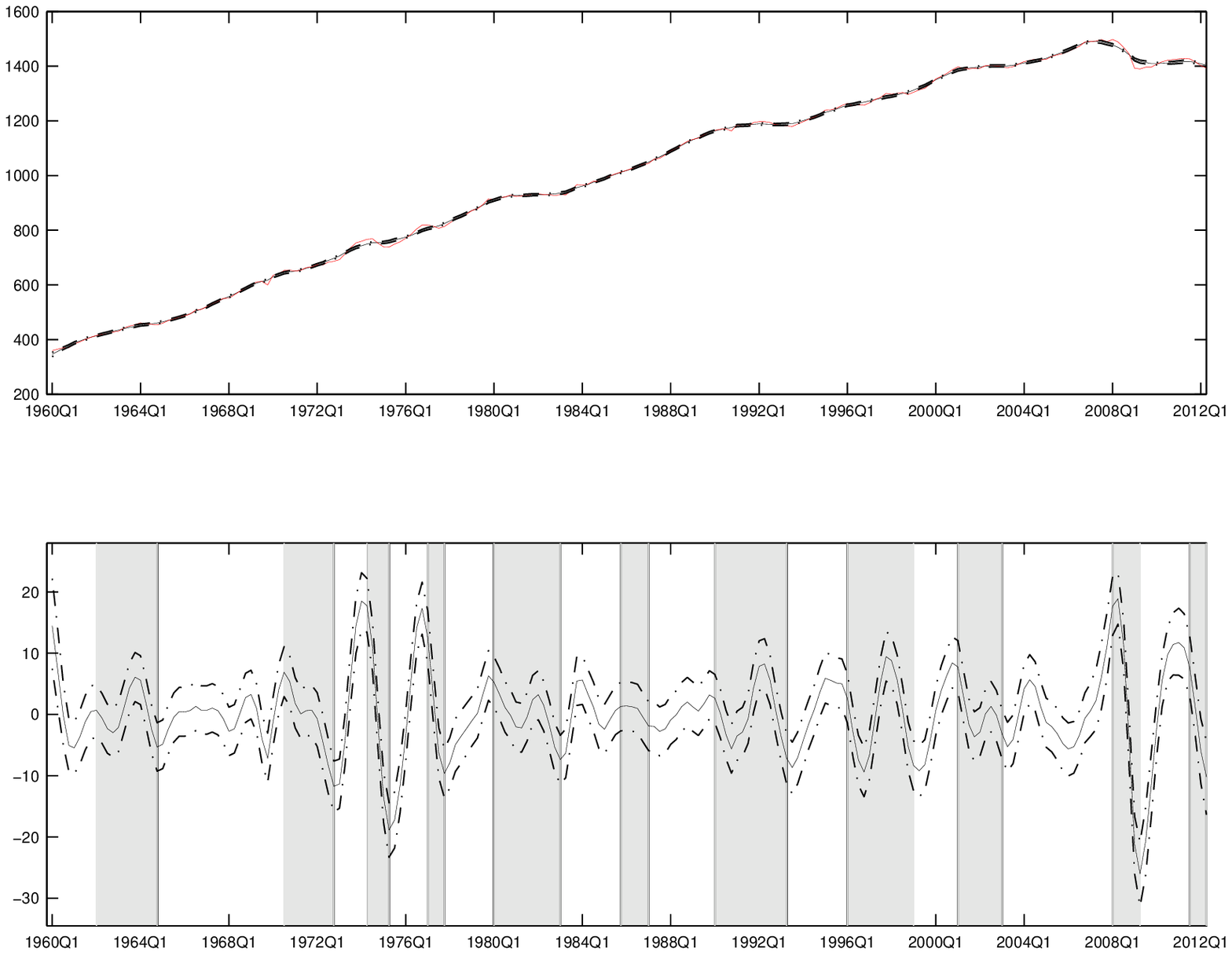}
\caption{\footnotesize{Posterior trend \textit{(upper panel)} and cycle \textit{(bottom panel)} estimates for the model $\mathcal{M}_{\sU}^{\sL\sT}$. The shaded area represents $95\%$ credible sets while the red line in the upper figure represents the real GDP.}}
\label{fig:univ_LTM_AR2_trend_cycle}
\end{center}
\end{figure}

\begin{figure}[!ht]
\begin{center}
\captionsetup{font={small}, labelfont=sc}
\includegraphics[width=1.0\linewidth]{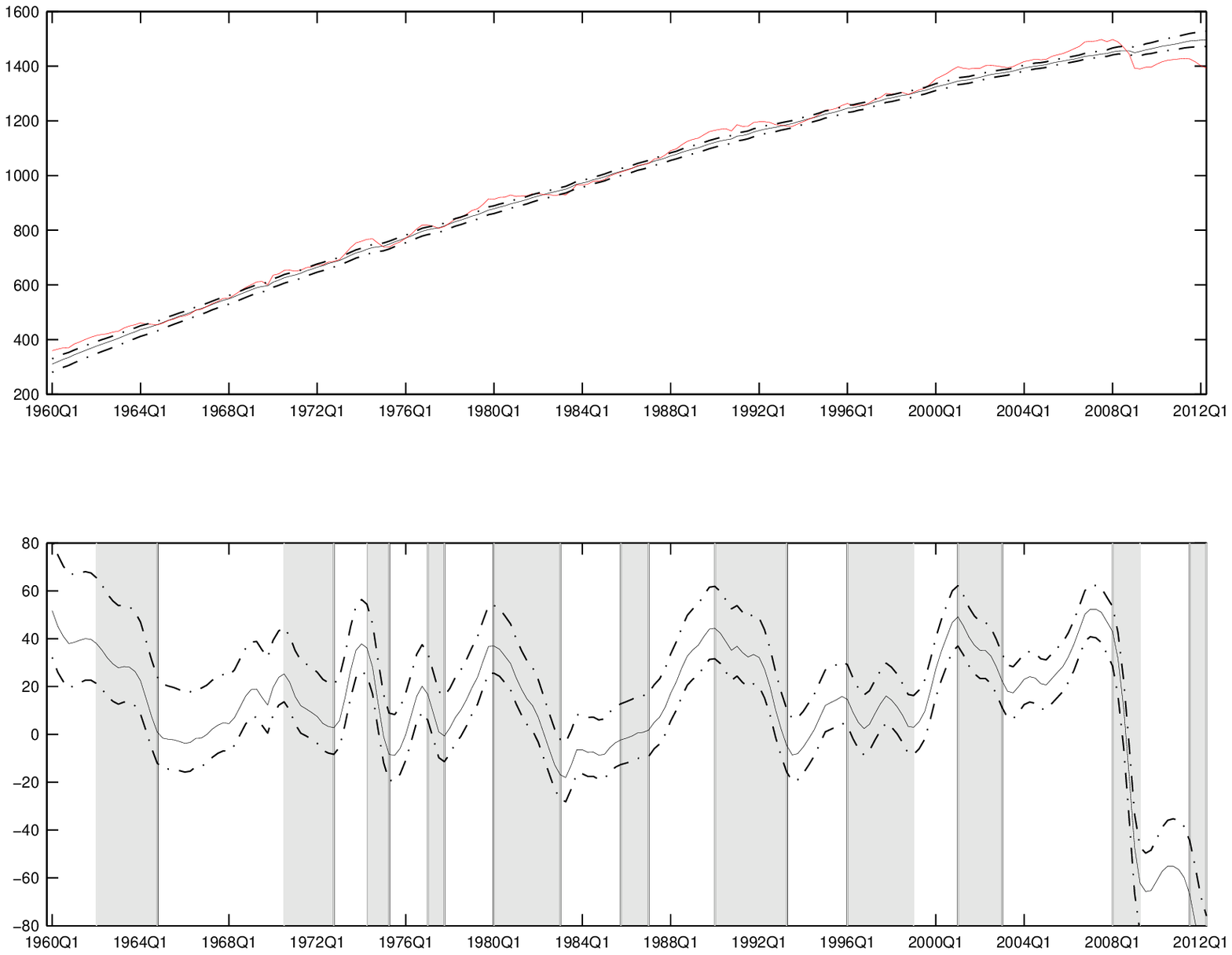}
\caption{\footnotesize{Posterior trend \textit{(upper panel)} and cycle \textit{(bottom panel)} estimates for the model $\mathcal{M}_{U}^{\sL\sL\sD}$. The shaded area represents $95\%$ credible sets while the red line in the upper figure represents the real GDP.}}
\label{fig:univ_LLD_AR2_trend_cycle}
\end{center}
\end{figure}

\begin{figure}[!ht]
\begin{center}
\captionsetup{font={small}, labelfont=sc}
\includegraphics[width=1.0\linewidth]{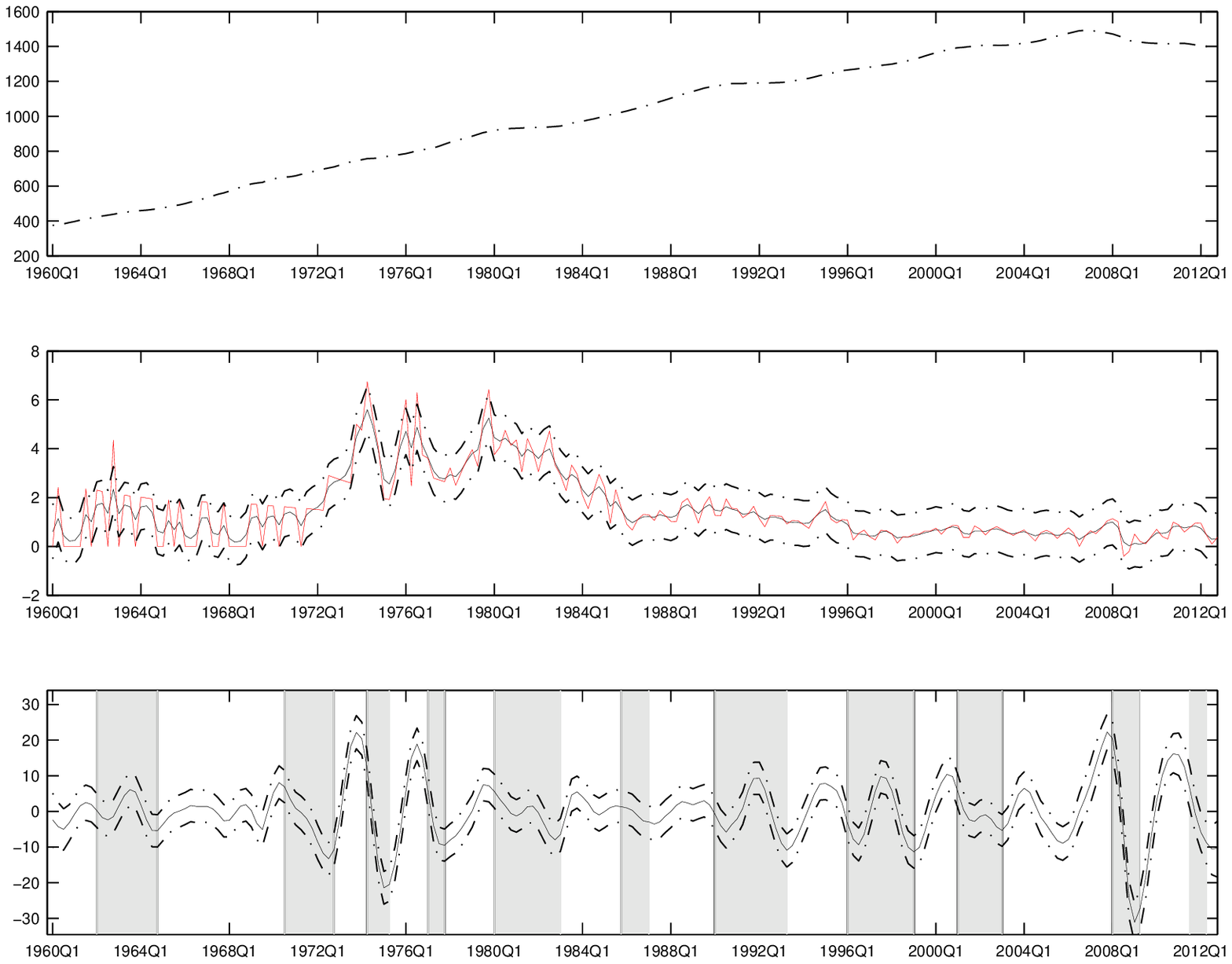}
\caption{\footnotesize{Posterior RGDP trend \textit{(upper panel)}, Inflation trend \textit{(middle panel)} and cycle \textit{(bottom panel)} estimates for the model $\mathcal{M}_{\sB}^{\sL\sT\sD}$. The shaded area represents $95\%$ credible sets while the red line in the upper figures represents the realised real GDP and inflation.}}
\label{fig:biv_OG_model_LTM_AR2_trend_cycle}
\end{center}
\end{figure}

%
\subsection{The data}
\label{sec:data}
%
\noindent We employ Italian time series from 1960:Q1 to 2013Q4 taken primarily from the OECD Quarterly National Accounts Database (\url{http://www.oecd.org/std/qna}). National accounts data for OECD member countries are based on each country's own System of National Accounts (SNA), 1993. Based on the data provided by each country, the OECD Secretariat has made quarterly estimates of the main expenditure components of GDP in order to publish historical data from 1960:Q1 when possible. The current systems of accounts have been linked to older systems based on different methodology in order to obtain longer time series. The method used by the Secretariat to link two time series from two different methodologies is as follows: for each individual series, the ratio between the new methodology data and the old methodology data in the first common year is calculated. This ratio is then multiplied by old methodology series for the time period that data have not been provided. The same method is applied to both current and volume estimates data.\newline 
\indent Output is the logarithm of annualised seasonally adjusted real GDP in OECD Base Year (2005) Euro. Following Kuttner \citeyearpar{kuttner.1994} and Planas \textit{et al.} \citeyearpar{planas_etal.2008}, the quarterly inflation was computed using the seasonally adjusted CPI (all items) data and was annualised. Figure 1 provides the graph of the actual inflation series from CPI--U along with the logarithm of the real GDP.
%
\subsection{Estimation results}
\label{sec:estimation_results}
%
In what follows we provide the Bayesian empirical analysis for the Output Gap models stated in Section \ref{sec:og_models}. In order to implement the inference we use the hyper--parameters values for each prior distribution defined in Section \ref{sec:prior_specification}. We obtain draws from the joint posterior distribution of the parameters and latent states following the Adaptive--Independent Metropolis--within--Gibbs algorithm detailed in Section \ref{sec:og_inference} for 200,000 times, with a burn-in phase of 100,000 iterations. Tables \ref{tab:univ_OG_models_param_est} an \ref{tab:biv_OG_models_param_est} in Appendix \ref{sec:appendix_B} report the summary statistics of the posterior MCMC draws for the univariate and bivariate OG models parameters. In particular, for all the considered model specifications, tables \ref{tab:univ_OG_models_param_est} and \ref{tab:biv_OG_models_param_est} report the mean, the standard deviation, the Maximum a Posteriori (MaP) as well as the $\text{HPD}_{95\%}$ credible sets evaluated over the post burn--in MCMC draws. To check the MCMC convergence we also calculate the Geweke's convergence diagnostics (see e.g. Geweke \citeyear{geweke.1992}, \citeyear{geweke.2005}) reported in the last column of each table which suggest that the convergence has been achieved for all the model parameters and specifications. Figures \ref{fig:univ_OG_model_LTM_post_draws} and \ref{fig:biv_OG_model_LTM_post_draws_part_1}--\ref{fig:biv_OG_model_LTM_post_draws_part_2} provide graphical representation of the MCMC output for the models $\mathcal{M}_{\sU}^{\sL\sT}$ and $\mathcal{M}_{\sB}^{\sL\sT}$. Visual inspection of the MCMC track (first column) as well as the autocorrelation between draws (last columns) reveal that convergence has been achieved. In particular, the autocorrelations of draws are quite low for the parameters of all the model specifications. The middle column's plots depict the histogram of the posterior draws for each parameters along with the corresponding prior density (red line). For all the models and models parameters, it is evident that priors are sufficiently diffuse when compared to the posterior histograms. We now turn to the analysis of the results.\newline 
\indent We start the analysis of the posterior estimate we got for the different models, by examining the parameters governing the cycle $\left(\rho,\lambda,\sigma_\kappa^2\right)$. Then, we move to the real GDP observation and trend innovations $\left(\sigma_\epsilon^2,\sigma_\eta^2\right)$. These parameters are common to all the model specifications. Finally, we analyse the output conceding the remaining parameters: $\sigma_\zeta^2$, the slope variance, and the parameters related to the bivariate specification, the inflation measurement and trend variances $\left(\sigma_\varepsilon^2,\sigma_\xi^2\right)$, and the OG loadings $\left(\theta_0,\theta_1\right)$.\newline
\indent Concerning the most important parameter governing the output gap, the cycles frequency $\lambda$ it can be seen from Table \ref{tab:univ_OG_models_param_est} that, concerning univariate models, the estimated frequency posterior mode ranges between about $0.43$ for the local level model $\mathcal{M}_{\sU}^{\sL\sL}$ and $0.54$ for the model $\mathcal{M}_{\sU}^{\sL\sT}$. The estimated frequency is much smaller for the local level model with fixed drift $\mathcal{M}_{\sU}^{\sL\sL\sD}$ because it correspond to a smoother potential estimate which does not adapt enough to the real output path. The periodicity posterior mode implied by those estimate is of about $3$ years for the $\mathcal{M}_{\sU}^{\sL\sL\sD}$ model and $3$ and an half years for the $\mathcal{M}_{\sU}^{\sL\sT}$ models. For both the univariate and bivariate models we observe that the posterior mode is a bit lager than the prior and the posterior variance is about ten time smaller when compared to the prior standard deviation. The bivariate (real GDP, inflation) models output reveals estimate of the posterior periodicity in line with those provided by simpler univariate models for the real GDP alone. For example, the cycle frequency posterior mode is about 3 years for the $\mathcal{M}_{\sB}^{\sL\sT}$ model with a similar dispersion. We conclude that the cycle period for the Italian economic cycle is about of three years, a result which is in line with previous findings of Zizza \citeyearpar{zizza.2006}. Moreover, comparing univariate and bivariate models we observe similar results both in term of cycle periodicity and its variance. Concerning the parameter that governs the cycle amplitude, $\rho$, we observe that the posterior mode estimate is considerably larger than the prior for all the model specifications we considered. Of course there are some differences across specifications which essentially reflect the imposed stochastic nature of the trend, but in every case our posterior estimate is sensible larger than that obtained in previous studies, see for example Zizza \citeyearpar{zizza.2006}. This suggests that demand shocks have a huge impact on real output and that their impact tends persist longer than for example in the US economy where Planas \textit{et al.} \citeyearpar{planas_etal.2008} found a coefficient equal to $0.82$. Moreover, comparing our estimate with that obtained for the US by Planas Planas \textit{et al.} \citeyearpar{planas_etal.2008}, we observe also that our estimate presents a level of uncertainty, measured by the standard deviation, which is sensible lower. The last parameter governing the cycle is its variance $\sigma^2_\kappa$. The posterior estimates we obtain for this parameter are quite similar across models in both locations and variances. The only exception is given by the model $\mathcal{M}_{\sU}^{\sL\sL\sD}$ where we found a posterior MaP which is three time larger on average than those obtained for all the other model specifications. Once again, the reason for this discrepancy should be looked for in the trend specification which result in a reduced flexibility in adapting to changes in the trend GDP. Unfortunately, prior studies on the italian business cycle do not report the estimated cycle variance, so we can not compare our result with those obtained using different time spanning or data.\newline
%
%
\begin{figure}[!t]
\begin{center}
\captionsetup{font={small}, labelfont=sc}
\includegraphics[width=0.65\linewidth,height=0.4\linewidth]{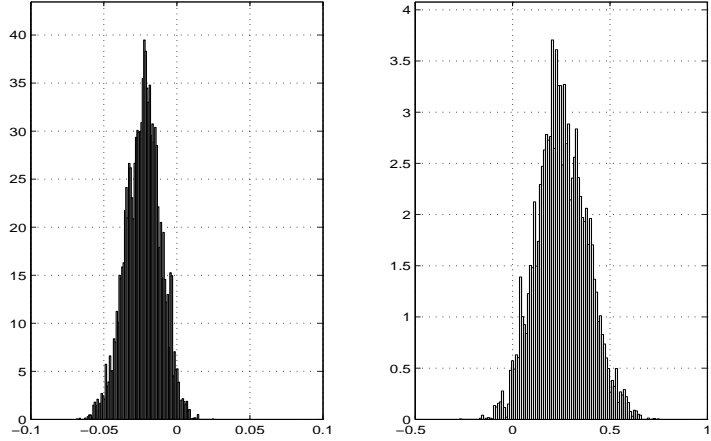}
\caption{\footnotesize{Posterior draws (post burn--in) of the change effect $\theta_0\phi_2$ \textit{(left panel)} and the level effect $\theta_0\left(\phi_1+\phi_2\right)+\theta_1$ \textit{(right panel)} for model $\mathcal{M}_{\sB}^{\sL\sT\sM}$.}}
\label{fig:BivOGModel_LTM_AR2_IT_OECD_Test_OG_param}
\end{center}
\end{figure}
%
%
\begin{figure}[!t]
\begin{center}
\captionsetup{font={small}, labelfont=sc}
\includegraphics[width=1.0\linewidth,height=0.45\linewidth]{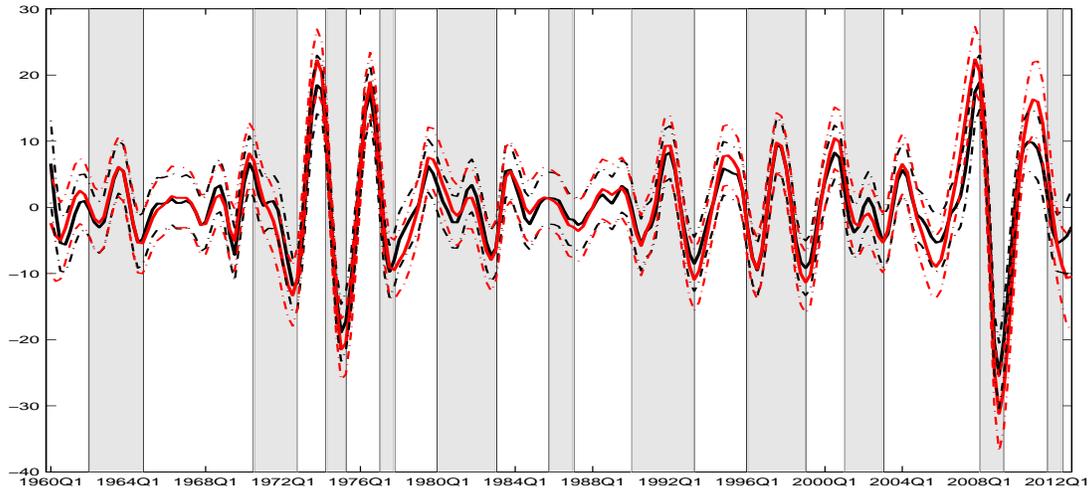}
\caption{\footnotesize{Comparison of the output gap estimates (MaP) implied by univariate $\mathcal{M}_{\sU}^{\sL\sT\sM}$ \textit{(black line)} and bivariate $\mathcal{M}_{\sB}^{\sL\sT\sM}$ \textit{(red line)} models. Dotted lines represent $95\%$ HPD credible sets.}}
\label{fig:OGModel_LTM_AR2_IT_OECD_Cycle_comparison}
\end{center}
\end{figure}
%
\indent The extent to which output gap influences the inflation dynamics in the long--run is measured by $\theta_0\left(\phi_1+\phi_2\right)+\theta_1$. As expected, we find that shocks on output gap have a strong positive effect on the first differences of inflation $\Delta^2p_t$. In particular we find that, on average, a 1\% deviation of the output from its long--run path produces an acceleration of the inflation growth for Italy of about $0.25\%$ with a standard deviation of $0.12\%$. Table \ref{tab:biv_OG_models_param_est} presents the parameter estimate of the OG loadings. We notice that, the second loading parameter $\theta_0\phi_2$ is negative by construction, because, as expected, the instantaneous correlation between the output gap $\kappa_t$ and inflation trend innovation $\xi$, measured by $\theta_0$, is positive, while the autoregressive parameter $\phi_2$ is imposed to be negative in order to have a pseudo--cyclical behaviour. Moreover, as observed by Proietti \citeyearpar{proietti.2009}, the OG loading polynomial $\theta_\psi\left(L\right)=\theta_{\psi,0}+\theta_{\psi,1}L$ can be decomposed into a permanent and a transitory component $\theta_{\psi}\left(L\right)=\theta_{\psi}\left(1\right)-\theta_{\psi,1}\Delta$, which enables us to isolate the level effect of the gap from the change effect. This latter effect, which is in general expected to be positive, turns out to be positive by construction in our model, as long as $\theta_0$ is positive. Finally, Figure \ref{fig:BivOGModel_LTM_AR2_IT_OECD_Test_OG_param} shows the posterior distribution of the change effect $\theta_0\phi_2$ and the level effect $\theta_0\left(\phi_1+\phi_2\right)+\theta_1$. The $95\%$ HPD of the former is $\left(0.0366,0.4999\right)$ while that for the latter is $\left(0.0344,0.4595\right)$ which confirms that the level effect of the output gap on the changes of inflation is significantly different from zero.\newline
\indent Concerning the coherence between the estimated cycles Figure \ref{fig:OGModel_LTM_AR2_IT_OECD_Cycle_comparison} report the OG obtained by fitting univariate (black line) and bivariate (red line) LTM. Extracted cycles display similar patters and similar $95\%$ HPD credible sets (dotted lines). Moreover, we observe the correlation between the two cycles is of the order of $0.965$. This means that there is substantial agreement between the OG estimates across univariate and bivariate models. 
\subsection{Discussion of the results}
\label{sec:disc_results}
%
\noindent The different methods employed to extract the Italian business cycle are now extensively analysed with respect to the existing business cycle datings for the italian economy.\newline
\indent Our results are compared with generally agreed upon international statistics on the business cycle dating and indicators. Among them, the most important are: the Composite Leading Indicator delivered by OECD and the business cycle chronology delivered by the Italian Bureau of Statistics (ISTAT, 2010). In what follows, we mainly refer to the OECD Indicator in order to be consistent with the data source. The leading indicator delivered by OECD consider monthly information coming from different sources such as the CPI and different measures of real production (see OECD System of Composite Leading Indicators, 2012; OECD Composite Leading Indicators: turning points of reference series and component series, 2014). The BC chronology provided by ISTAT instead consider the NBER methodology. For a comprehensive treatment of the different statistical methodologies of dating the business cycle phases with application to the Italian economy, we refer to Bruno and Otranto \citeyearpar{bruno_otranto.2004}.\newline 
%
%
%
\indent Our estimates of the italian output gap turn out to be quite  consistent with the italian business cycle dating indicators currently available. In Figure \ref{fig:OGModel_LTM_AR2_IT_OECD_Cycle_comparison}, univariate and bivariate OG models with local linear trend specification of the GDP dynamics, i.e. $\mathcal{M}_\sU^{\sL\sT\sM}$ and $\mathcal{M}_\sB^{\sL\sT\sM}$ are compared with the OECD business cycle turning points' reference chronology. By a visual inspection it seems that our estimated gaps perform very well to capture the major turning points, in particular starting from the seventies onward. It should be noted in advance, that some small discrepancies between the signal provided by our models and the OECD chronology should be imputed to the different frequency of the data. In particular, we consider quarterly data, while the OECD makes use of monthly data. However, some mismatches are easily detectable. For instance, from 1962 to 1965, we find evidence of a cycle completion signal rather than a contraction phase as detected by the OECD dating system. This evidence could be due to the fact that the OECD dating methodology employs the monthly industrial production as the reference series to construct its indicator, rather than the real quarterly GDP, as it is in our case. In fact, the industrial production series turned out to be affected by the large investments drop right after the wage shock occurred in 1962, while the Italian economy as a whole was experiencing the first inflationary pressure and was approaching to the full capacity of utilisation. To be exhaustive, we stress that the unreported ISTAT chronology better adapts to our output gap pattern in that period.\newline
\indent Another OG milder fluctuation is registered within the period 1968--1970 which was characterised by a massive series of strikes in the factories, reaching its peak in the so--called \qmo hot autumn\qmc. The large amplitude of the oscillations registered in the 70's can be partly attributed  to the fall of the Bretton Woods Agreement  the consequent devaluation of the Italian currency, started on February 1972, on one hand, and to the first oil shock of October 1973, on the other hand.\newline
\indent The 1980's open with the beginning of a recession phase which lasted 3 years. Those years were mainly characterised by a new raise in the inflation levels, while the industrial production index stayed put. Our estimates perfectly captures  the upturn of 1983, in which production and investments started to raise again triggering a strong boost in the industrial productivity. The growth phase, more moderate with respect to the one experienced during the period 1976--1979 (as it is easily detectable from a visual inspection of our graph), lasted until 1989, except for a one--year mild contraction in 1986. That phase reached its peak in 1988, in which GDP grew of $4,1\%$ (see Battilossi \citeyearpar{battilossi.1999}).\newline
\indent The 1990's economic downturn follows the downward economic performance of the world economy, according to the OECD business cycle indicator. However, our estimated gap seems to be more consistent with the ISTAT chronology which detects the downturn peak in April 1992.During that year,  Italy experienced a severe currency crisis and a dramatic change in the budget policy in order to comply with the requirements of the Maastricht treaty. The following year was characterised by a drop in investment and industrial production and by a persistent political instability and it represents the trough of the previous economic cycle.
The recovery period of 1994--1995 was mainly due to the foreign demand. During that period, the Italian exports benefited from the previous devaluation of the Lira, and they grew at a rate of $10\%$.
The Italian economy shrank in 1996, in which industrial production decreased of $2,9\%$, mainly because of the slowdown of exports which could not take advantage anymore of the devaluation of the Lira, which started a slow revaluation against the other European currencies. 
According to our estimates, the OG becomes positive in 1997, when the industrial production 
experienced an acceleration during the second half of the year. Industrial profits were the highest of that decade, mainly because of a big boost in the labour productivity. It is evident from our graphs that the 1990's were characterised by a cycle period being on average lower than the one of the previous decade and by an higher amplitude.\newline
Regarding the 2000s, our estimates are quite in line with  the OECD's dating.
The higher amplitude registered after the first five years of this period clearly reflects the financial crisis started in 2007, which led to the global recession of 2008-2012.

%
%
\indent From an historical perspective, another important result is the unambiguous role that output gap plays in influencing inflation. Our findings strongly suggest that OG has a positive net effect on inflation in the long run. In fact the pro--cyclicality of inflation was quite an expected result. The period under investigation has been characterised from its beginning by the progressive tendency of the Italian economy towards full employment. In that circumstances, the typical Phillips curve trade--off arises and the price dynamics finds to have one of its own explanations in the excess demand in goods and services markets which causes inflationary pressures on the real economy. 
%
%
\section{Conclusion}
\label{sec:conclusion}
%
\noindent In this article we develop Bayesian methods to extract the Italian OG for the period 1960--Q1 to 2013--Q2. This methodology enables potential output and OG estimates to incorporate prior information. The intrinsically unobservable nature of the OG led it to be represented as a latent state linked to the observable processes using a state space models. The models we consider ecompass several univariate and bivariate specifications differing for the specified trend dynamics. Univariate specifications rely on the deseasonalised real GDP series recently released by the OECD and available at the OECD Quarterly National Accounts Database (\url{http://www.oecd.org/std/qna}), while bivariate models extend them introducing the price dynamics through a backward Phillips curve relation as in Kuttner \citeyearpar{kuttner.1994} and Planas \textit{et al.} \citeyearpar{planas_etal.2008}. As said throughout the paper, the real GDP series made recently available by the OECD, extends previous series and permits the analysis of the OG dynamics for the period 1960--1980.
In all the considered models we specify the OG dynamics as a cyclical AR(2) process reparameterised in terms of the polar coordinates of the characteristic equation roots, as in Planas \textit{et al.} \citeyearpar{planas_etal.2008}.\newline
%
\indent The statistical analysis involves a new Adaptive--Independent Metropolis--Hastings--within--Gibbs (AIMH--GS) algorithm to simulate from the high dimension posterior parameter space. Adaptive MCMC methods have been proved to be successful in sampling from intractable and multimodal distribution, simply by iteratively adapting the proposal to the target using past draws. The adaptive Metropolis method is particularly effective for sampling periodicity and amplitude parameters which enters the likelihood function in a highly non--linear and complicated way involving both the location and scale parameters of the respective full conditional distributions.\newline
\indent Our results confirm that the Italian data agree with Phillips curve theory. In particular, inflation plays a role in characterising the business cycle over the different historical phases. This essentially makes the extracted Italian OG an important indicator of inflation pressures on the real side of the economy as suggested by the Phillips theory.
Moreover, our estimate of the sequence of peaks and troughs of the Output Gap is in line with the OECD official dating of the Italian business cycle.
%
%
%
%
However, providing an interpretation of the underlying mechanism governing the business cycle pattern is not an easy task, and it is far beyond the purpose of this paper. Adopting a unifying conceptual framework to interpret every business cycle fluctuation is a naive attempt. Instead, it would be much more appropriate to consider every cycle as an historical episode, embedded in a specific economic contest. We firmly believe that, in order to obtain an explanation of the business cycle fluctuations, our aggregate analysis is just the first step and a disaggregate analysis at a sectorial level would be necessary to address the causes of each fluctuation. As Gallegati and Stanca \citeyearpar{gallegati_stanca.1998} pointed out it is difficult, to search for a causal explanation when microeconomic relations influence the aggregates.\newline 
\indent The analysis carried out in this paper can be extended in several directions. It would be interesting, for example, to split the sample in order to see the difference in the impact upon inflation in different sub periods to check the stability of the estimate of the cycle parameters such as phase and amplitude. Another contribution could be to check the robustness of the result with respect to the choice of the series. Several price indicators, such as the GDP deflator or price indexes, such as wholesale or producer prices, can be compared. On the real side, additional information on the real business cycle can be provided by the industrial production index which is available on a monthly basis. The availability of data at different frequencies opens the opportunity of providing a monthly trend--cycle decomposition which is of particular interest as leading indicator of the economic activity. Furthermore, given the growing integration of the economies of the last decades, another important improvement would be to jointly model data of different economies. We will leave these issues for future research.\newline  
\indent Given the small amount of macroeconomic knowledge and business
cycle studies available for the Italian case, we believe that this Bayesian analysis will be appealing to practitioners.
%
\section*{Acknowledgments}
%
\noindent This research is supported by the Italian Ministry of Research PRIN 2013--2015, ``Multivariate Statistical Methods for Risk Assessment'' (MISURA), and by the ``Carlo Giannini Research Fellowship'', the ``Centro Interuniversitario di Econometria'' (CIdE) and ``UniCredit Foundation''. The authors are very grateful to Riccardo Sucapare for his friendly and active collaboration. 
\clearpage
\newpage
%

\appendix
%
\section{MCMC output}
\label{sec:appendix_A}
%

\begin{figure}[!ht]
\begin{center}
\captionsetup{font={small}, labelfont=sc}
\includegraphics[width=1.0\linewidth]{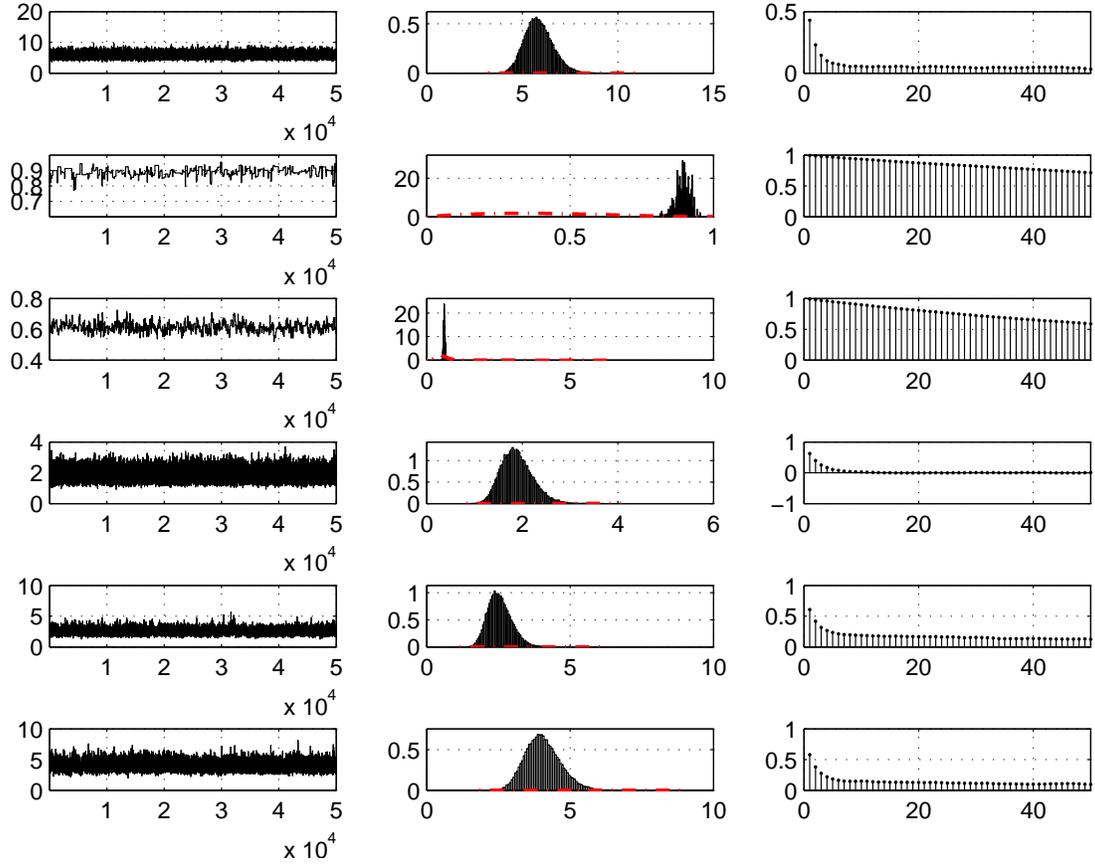}
\caption{\footnotesize{Posterior draws (post burn--in) from the univariate output gap model parameters. Posterior track \textit{(left panel)}, posterior histogram \textit{(middle panel)}, autocorrelation of the posterior draws, \textit{(right panel)}. The red dotted line denotes the prior density.}}
\label{fig:univ_OG_model_LTM_post_draws}
\end{center}
\end{figure}

\begin{figure}[!ht]
\begin{center}
\captionsetup{font={small}, labelfont=sc}
\includegraphics[width=1.0\linewidth]{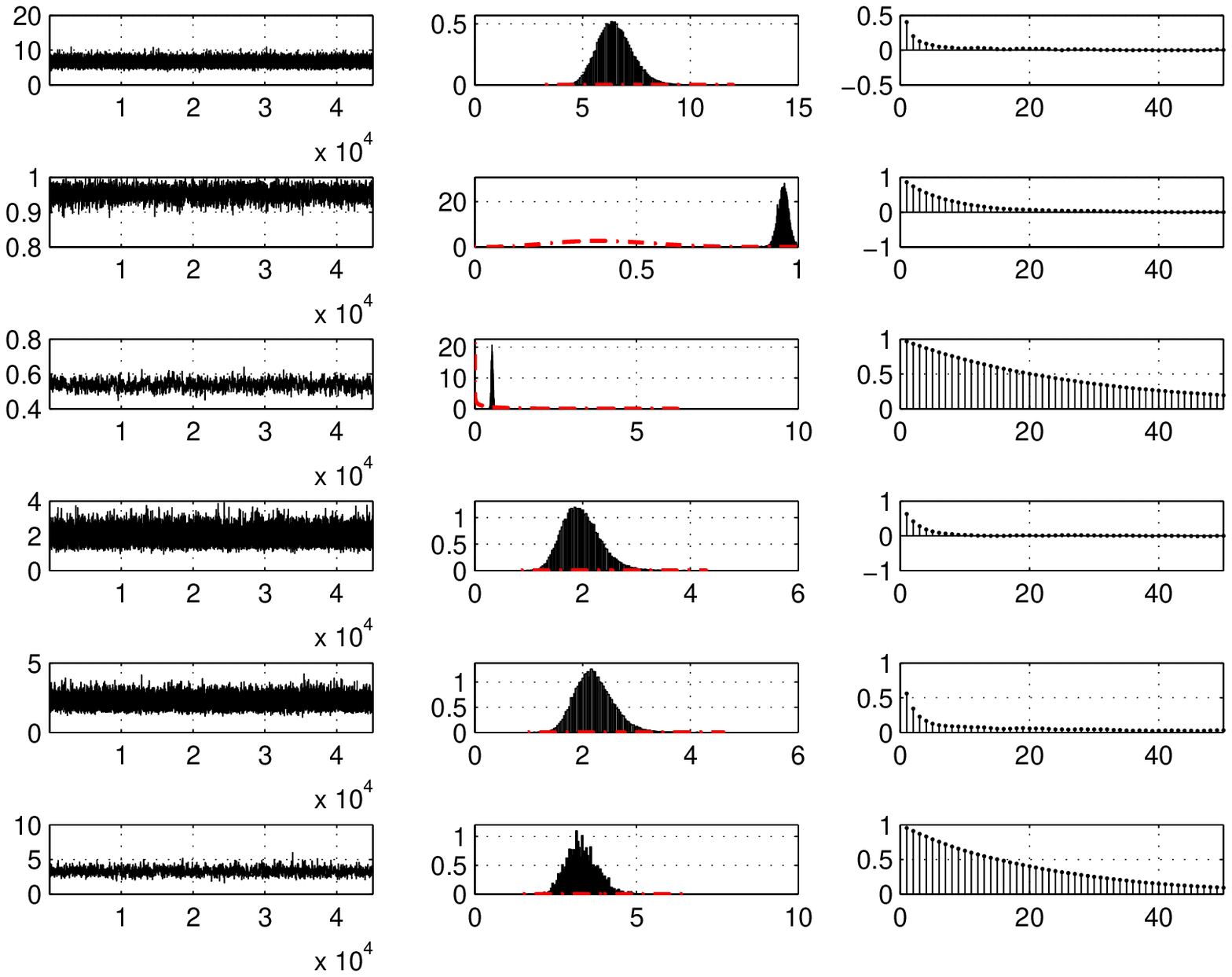}
\caption{\footnotesize{Posterior draws (post burn--in) from the bivariate output gap $\mathcal{M}_{\sB}^{\sL\sT}$ model parameters. Posterior track \textit{(left panel)}, posterior histogram \textit{(middle panel)}, autocorrelation of the posterior draws, \textit{(right panel)}. The red dotted line denotes the prior density.}}
\label{fig:biv_OG_model_LTM_post_draws_part_1}
\end{center}
\end{figure}

\begin{figure}[!ht]
\begin{center}
\captionsetup{font={small}, labelfont=sc}
\includegraphics[width=1.0\linewidth]{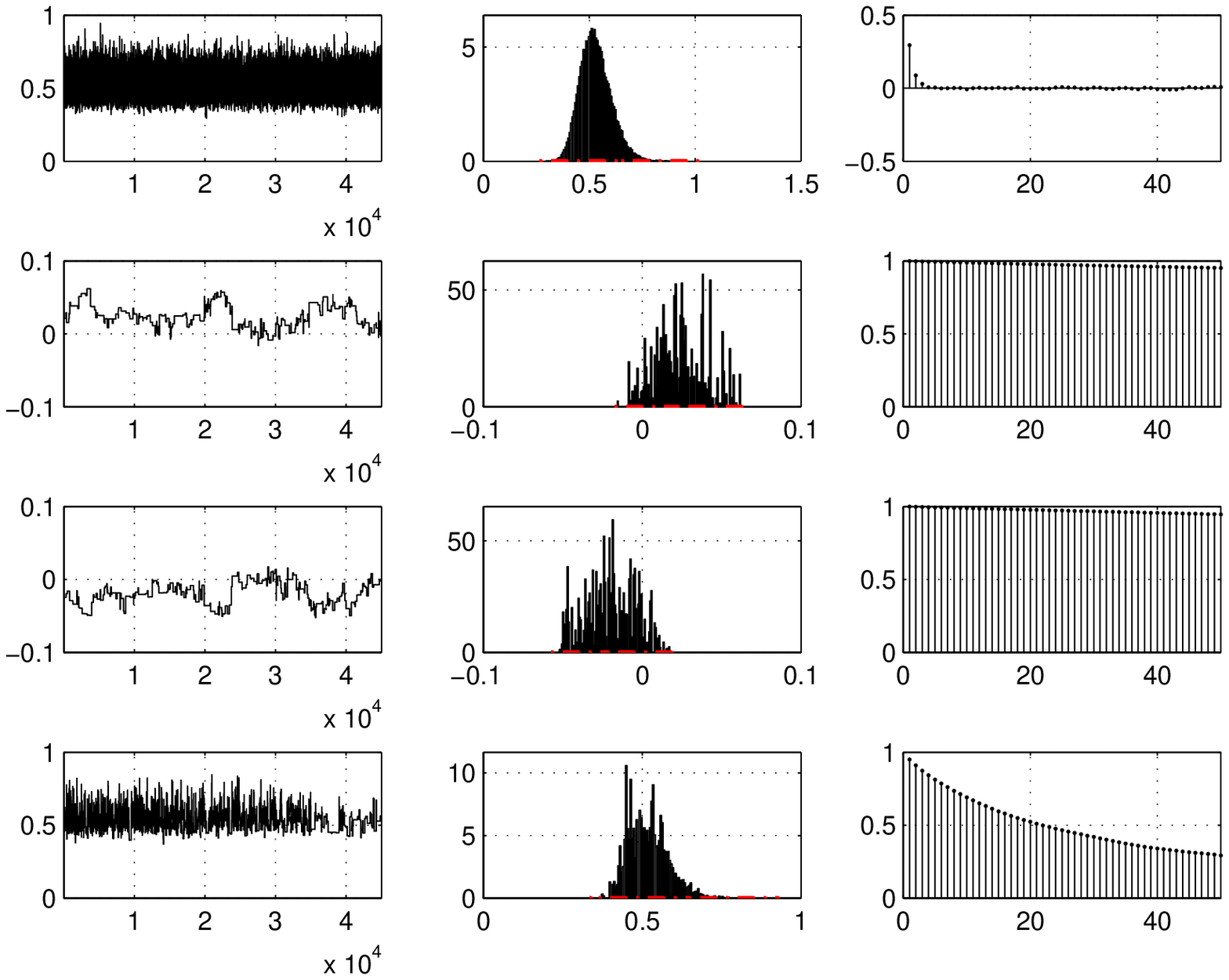}
\caption{\footnotesize{Posterior draws (post burn--in) from the bivariate output gap $\mathcal{M}_{\sB}^{\sL\sT}$ model parameters. Posterior track \textit{(left panel)}, posterior histogram \textit{(middle panel)}, autocorrelation of the posterior draws, \textit{(right panel)}. The red dotted line denotes the prior density.}}
\label{fig:biv_OG_model_LTM_post_draws_part_2}
\end{center}
\end{figure}

\begin{table}[!h]
\captionsetup{font={footnotesize}, labelfont=sc}
\begin{small}
\begin{center}
\tabcolsep=2.0mm
    \begin{tabular}{ccccccc}
     \toprule
    & \multicolumn{6}{c}{$\mathcal{M}_\sU^{\sL\sL}$} \\
\cmidrule(lr){2-7}
\multirow{2}{*}{Param}& \multicolumn{1}{c}{\multirow{2}{*}{Mean}}  & \multirow{2}{*}{Std. Dev.} & \multirow{2}{*}{MaP} & \multicolumn{2}{c}{HPD$_{95\%}$} & \multicolumn{1}{c}{\multirow{2}{*}{Geweke}}\\
 \cmidrule(lr){5-6}  
     & & &  & Lower & Upper &         \\
     \cmidrule(lr){1-1}\cmidrule(lr){2-7}
$\sigma_{\epsilon}^2$&2.6758  &0.4154  &2.2272  &1.8951  &3.4953  &-0.5617  \\
$\rho$&0.9450  &0.0147  &0.9389  &0.8962  &0.9736  &1.1875  \\
$\lambda$&0.3938  &0.0271  &0.4372  &0.3397  &0.4465  &1.5533  \\
$\sigma_{\eta}^2$&50.4346  &5.2176  &74.5991  &40.6264  &60.9359  &1.2301  \\
$\sigma_{\kappa}^2$&3.5024  &0.5272  &2.9663  &2.5176  &4.5519  &-1.2025  \\
\hline
& \multicolumn{6}{c}{$\mathcal{M}_\sU^{\sL\sT}$} \\
\cmidrule(lr){2-7}
\multirow{2}{*}{Param}& \multicolumn{1}{c}{\multirow{2}{*}{Mean}}  & \multirow{2}{*}{Std. Dev.} & \multirow{2}{*}{MaP} & \multicolumn{2}{c}{HPD$_{95\%}$} & \multicolumn{1}{c}{\multirow{2}{*}{Geweke}}\\
 \cmidrule(lr){5-6}  
     & & &  & Lower & Upper &         \\
     \cmidrule(lr){1-1}\cmidrule(lr){2-7}
$\sigma_{\epsilon}^2$&5.8878  &0.7452  &8.8266  &4.4828  &7.3559  &1.2956  \\
$\rho$&0.8943  &0.0272  &0.8992  &0.8370  &0.9408  &1.1892  \\
$\lambda$&0.6145  &0.0286  &0.5400  &0.5564  &0.6717  &-0.6165  \\
$\sigma_{\eta}^2$&1.8777  &0.3280  &1.3823  &1.2739  &2.5336  &1.1395  \\
$\sigma_{\zeta}^2$&2.5499  &0.4203  &2.7753  &1.7743  &3.3952  &-1.3122  \\
$\sigma_{\kappa}^2$&4.0217  &0.6151  &4.7596  &2.8818  &5.2683  &-1.7832  \\
\hline
& \multicolumn{6}{c}{$\mathcal{M}_\sU^{\sL\sL\sD}$} \\
\cmidrule(lr){2-7}
\multirow{2}{*}{Param}& \multicolumn{1}{c}{\multirow{2}{*}{Mean}}  & \multirow{2}{*}{Std. Dev.} & \multirow{2}{*}{MaP} & \multicolumn{2}{c}{HPD$_{95\%}$} & \multicolumn{1}{c}{\multirow{2}{*}{Geweke}}\\
 \cmidrule(lr){5-6}  
     & & &  & Lower & Upper &         \\
     \cmidrule(lr){1-1}\cmidrule(lr){2-7}
$\sigma_{\epsilon}^2$&4.9518  &0.7131  &6.1168  &3.6094  &6.3695  &-0.1804  \\
$\rho$&0.7947  &0.0237  &0.8737  &0.7442  &0.8395  &-0.3322  \\
$\lambda$&0.0810  &0.0227  &0.0657  &0.0525  &0.1167  &-0.6068  \\
$\sigma_{\eta}^2$&6.7555  &1.3245  &5.5008  &4.3212  &9.3962  &-0.7198  \\
$\sigma_{\kappa}^2$&13.3922  &2.6218  &20.7430  &10.6911  &15.1906  &1.3280  \\
\hline
& \multicolumn{6}{c}{$\mathcal{M}_\sU^{\sI\sR\sW}$} \\
\cmidrule(lr){2-7}
\multirow{2}{*}{Param}& \multicolumn{1}{c}{\multirow{2}{*}{Mean}}  & \multirow{2}{*}{Std. Dev.} & \multirow{2}{*}{MaP} & \multicolumn{2}{c}{HPD$_{95\%}$} & \multicolumn{1}{c}{\multirow{2}{*}{Geweke}}\\
 \cmidrule(lr){5-6}  
     & & &  & Lower & Upper &         \\
     \cmidrule(lr){1-1}\cmidrule(lr){2-7}
$\sigma_{\epsilon}^2$&7.4904  &0.8745  &10.5952  &5.8423  &9.2310  &0.8068  \\
$\rho$&0.8998  &0.0250  &0.9071  &0.8487  &0.9469  &-0.0609  \\
$\lambda$&0.6291  &0.0274  &0.5673  &0.5760  &0.6822  &-1.1226  \\
$\sigma_{\zeta}^2$&2.9450  &0.4585  &3.0665  &2.0937  &3.8582  &-1.9009  \\
$\sigma_{\kappa}^2$&4.3272  &0.6607  &4.7758  &3.1105  &5.6411  &0.5562  \\
          \bottomrule
    \end{tabular}
\caption{\footnotesize{Summary statistics of the posterior MCMC draws for the Univariate OG model specifications. The fourth column denoted \qmo MaP\qmc, reports the Maximum a Posteriori estimate, while column 5 and 6 report the lower and upper bound for the $95\%$ high posterior credible sets. The last column, denoted \qmo Geweke\qmcsp reports the absolute value of the Geweke's (1995) convergence statistic.}}
  \label{tab:univ_OG_models_param_est}
    \end{center}
    \end{small}
\end{table}

\begin{table}[!h]
\captionsetup{font={footnotesize}, labelfont=sc}
\begin{small}
\begin{center}
\tabcolsep=2.0mm
    \begin{tabular}{ccccccc}
    \hline
    & \multicolumn{6}{c}{$\mathcal{M}_{\sB}^{\sL\sL}$} \\
\cmidrule(lr){2-7}
\multirow{2}{*}{Param}& \multicolumn{1}{c}{\multirow{2}{*}{Mean}}  & \multirow{2}{*}{Std. Dev.} & \multirow{2}{*}{MaP} & \multicolumn{2}{c}{HPD$_{95\%}$} & \multicolumn{1}{c}{\multirow{2}{*}{Geweke}}\\
%
 \cmidrule(lr){5-6}  
     & & &  & Lower & Upper &         \\
     \cmidrule(lr){1-1}\cmidrule(lr){2-7}
$\sigma_{\epsilon}^2$&6.5032  &0.8004  &9.9586  &4.9884  &8.0907  &-0.1671  \\
$\rho$&0.9536  &0.0159  &0.9642  &0.9217  &0.9842  &-1.2168  \\
$\lambda$&0.5369  &0.0241  &0.5558  &0.4880  &0.5837  &0.3445  \\
$\sigma_{\eta}^2$&1.9785  &0.3490  &1.8449  &1.3357  &2.6788  &-1.0410  \\
$\sigma_{\zeta}^2$&2.2136  &0.3456  &2.5643  &1.5629  &2.8973  &0.2063  \\
$\sigma_{\kappa}^2$&3.3077  &0.4767  &3.2400  &2.4130  &4.2214  &0.2542  \\
$\sigma_{\varepsilon}^2$&0.5295  &0.0729  &0.4390  &0.3928  &0.6752  &0.3881  \\
$\theta_0$&0.0241  &0.0160  &0.0385  &-0.0051  &0.0571  &9.6003  \\
$\theta_1$&-0.0198  &0.0155  &-0.0314  &-0.0497  &0.0075  &-9.8364  \\
$\sigma_{\xi}^2$&0.5184  &0.0631  &0.4254  &0.4096  &0.6460  &-1.7092  \\
$\theta_0\phi_1+\theta_1$&0.2767  &0.1415  &0.4903  &-0.0070  &0.5463  &0.7875  \\
$\theta_0\phi_2$&-0.0231  &0.0118  &-0.0388  &-0.0462  &-0.0001  &-0.6817  \\
\hline
    \end{tabular}
\caption{\footnotesize{Summary statistics of the posterior MCMC draws for the Bivariate OG model specifications. The fourth column denoted \qmo MaP\qmc, reports the Maximum a Posteriori estimate, while column 5 and 6 report the lower and upper bound for the $95\%$ high posterior credible sets. The last column, denoted \qmo Geweke\qmcsp reports the absolute value of the Geweke's (1995) convergence statistic.}}
  \label{tab:biv_OG_models_param_est}
    \end{center}
    \end{small}
\end{table}
%
%
\clearpage
\newpage


\end{document}